\documentclass[journal=jacsat,manuscript=article,layout=traditional]{achemso}

\usepackage[version=3]{mhchem} 

\usepackage{amssymb}
\usepackage{graphicx}% Include figure files
\usepackage{dcolumn}% Align table columns on decimal point
\usepackage{bm}% bold math
\usepackage{color}
\usepackage{ulem}

\author{Chiara Guidolin}
\affiliation[BIOMETRA]
{Dipartimento di Biotecnologie Mediche e Medicina Traslazionale, Università
degli Studi di Milano, Via. F.lli Cervi 93, Segrate (MI) I-20090,
Italy}
\author{Christopher Heim}
\affiliation{NanoTemper Technologies GmbH, Munich, Germany}
\author{Nathan B P Adams}
\affiliation{NanoTemper Technologies GmbH, Munich, Germany}
\author{Philipp Baaske}
\affiliation{NanoTemper Technologies GmbH, Munich, Germany}
\author{Valeria Rondelli}
\affiliation[BIOMETRA]
{Dipartimento di Biotecnologie Mediche e Medicina Traslazionale, Università
degli Studi di Milano, Via. F.lli Cervi 93, Segrate (MI) I-20090,
Italy}
\author{Roberto Cerbino}
\affiliation{Faculty of Physics, University of Vienna, Boltzmanngasse 5, Vienna 1090, Austria}
\email{roberto.cerbino@univie.ac.at}
\author{Fabio Giavazzi}
\email{fabio.giavazzi@unimi.it}
\affiliation[BIOMETRA]
{Dipartimento di Biotecnologie Mediche e Medicina Traslazionale, Università
degli Studi di Milano, Via. F.lli Cervi 93, Segrate (MI) I-20090,
Italy}
\title[]
  {Protein sizing with Differential Dynamic Microscopy}% \footnote{A footnote for the title}}

\definecolor{FG}{rgb}{1,0,0}

\definecolor{CG}{rgb}{0,0,1}

\def\REV#1{\textcolor{black}{#1}}

\begin{document}

\maketitle

%\begin{tocentry}
%Inside the \texttt{tocentry} environment, the font used is Helvetica 8 pt, as required by \emph{Journal of the American Chemical Society}.
%\end{tocentry}

\begin{abstract}
Introduced more than fifty years ago, dynamic light scattering is routinely used to determine the size distribution of colloidal suspensions, as well as of macromolecules in solution, such as proteins, nucleic acids, and their complexes. More recently, differential dynamic microscopy has been proposed as a way to perform dynamic light scattering experiments with a microscope, with much less stringent constraints in terms of cleanliness of the optical surfaces, but a potentially lower sensitivity due to the use of camera-based detectors. In this work, we push bright-field differential dynamic microscopy beyond known limits and show it to be sufficiently sensitive to size small macromolecules in diluted solutions. By considering solutions of three different proteins (Bovine Serum Albumin, Lysozyme, and Pepsin), we accurately determine the diffusion coefficient and hydrodynamic radius of both single proteins and small protein aggregates down to concentrations of a few milligrams per milliliter. In addition, we present preliminary results showing unexplored potential for the determination of virial coefficients. Our results are in excellent agreement with the ones obtained in parallel with a state-of-the-art commercial dynamic light scattering setup, showing that differential dynamic microscopy represents a valuable alternative for rapid, label-free protein sizing with an optical microscope. 

\end{abstract}

\section{Introduction}

Dynamic light scattering (DLS) is an extremely sensitive optical method to probe the dynamics of complex fluids: coherent light with wavelength $\lambda$ illuminates the sample and the temporally fluctuating scattering intensity at an angle $\theta$ is quantitatively analyzed to extract the lifetime of the refractive index fluctuations occurring in the sample at the length-scale $L\simeq \lambda/2\sin(\theta/2)$ \cite{pecora1964doppler,debye1965spectral}. For a diluted suspension of small particles, the refractive index distribution is mainly determined by the particle concentration, which fluctuates over time due to Brownian motion. In this case, measuring the correlation time of the intensity fluctuations enables estimating the diffusion coefficient of the particles from which, if the viscosity of the solvent is known, their hydrodynamic radius can be determined \textit{via} the Stokes-Einstein relation \cite{einstein1956investigations}.
Firstly demonstrated in the late sixties
%of the twentieth century
as a rapid and non-invasive method to measure the size of proteins, nucleic acids, viruses, and synthetic nanoparticles in solution\cite{dubin1967observation, foord1970determination}, DLS rapidly grew in popularity, and nowadays commercial DLS instruments are part of the standard instrumentation of many pharmacology, physics, biology, and chemistry laboratories, both in academic and in industrial contexts.
Since its introduction, one of the key applications of DLS has been the determination of the size of  macromolecules and supramolecular structures in a range roughly comprised between a fraction of a nanometer and a few microns \cite{stetefeld2016dynamic}.
Moreover, the strong dependence of the scattered intensity $I$ on the particle size (for small particles $I$ scales as the sixth power of the radius\cite{hulst1981light}) makes DLS particularly effective in detecting the presence of aggregates and monitoring aggregation processes \cite{stetefeld2016dynamic}.

A downside of the excellent sensitivity of DLS is its extreme susceptibility to the presence of impurities both in the sample and along the optical path, implying meticulous care in sample preparation (\textit{e.g.} the need for multiple filtration steps) and high cleanliness standards for all the optical surfaces \cite{stetefeld2016dynamic}. Moreover, due to its strong susceptibility to multiple scattering, DLS is suitable for probing only fairly transparent solutions.
This strongly narrows the range of concentrations that DLS can probe and prevents its application to even slightly turbid materials or in harsh experimental conditions. 
To overcome these limitations, different variants have been proposed over the years, aimed either at reducing the effect of multiple scattering, for example exploiting different cross-correlation schemes, as in two-colour and three-dimensional dynamic light scattering\cite{pusey1999suppression}, or at extracting dynamical information from the multiply-scattered light, as in diffusing wave spectroscopy (DWS)\cite{PhysRevLett.60.1134}.
However, except for DWS, which operates in the strong multiple scattering limit and is suitable for highly turbid materials, none of the aforementioned alternatives to DLS has achieved comparable commercial success, mainly due to the increased operational and instrumental complexity compared to standard DLS.

In 2008, Cerbino and Trappe demonstrated that a conventional optical microscope using white-light illumination could be used to extract the same information obtained with DLS, by analyzing temporal sequences of images of the sample acquired in direct space \cite{cerbino2008differential}. This approach, named differential dynamic microscopy (DDM), relies on a fully automated and operator-independent procedure that calculates simultaneously for different wavevectors $q$ the temporal correlations of the spatially-Fourier-transformed sample images\cite{giavazzi2009scattering}. 
DDM has been successfully demonstrated in a wide range of applications including the characterization of the Brownian dynamics of diluted \cite{cerbino2008differential, bayles2016dark} and concentrated \cite{lu2012characterizing,brizioli2022reciprocal} colloidal suspensions, the microrheology of complex fluids \cite{bayles2017probe, edera2017differential, cerbino2022differential}, the active motion of swimming microorganisms \cite{wilson2011differential, martinez2012differential, lu2012characterizing} and crawling cells \cite{giavazzi2018tracking},  the protein absorption on functionalized  nanoparticles\cite{latreille2022situ}, and the diffusive behavior of large protein clusters \cite{safari2015differential}. 

The close formal correspondence with DLS enabled expanding the range of applicability of DDM by directly exploiting or adapting both well-established experimental geometries (like depolarized \cite{giavazzi2014viscoelasticity, giavazzi2016simultaneous} or wide-angle scattering \cite{cerbino2017dark} to measure the rototranslational dynamics of anisotropic particles)  and analytical tools (like cumulant expansions \cite{safari2015differential,giavazzi2016simultaneous} to estimate polydispersity or even more refined inversion schemes, like CONTIN\cite{provencher1982contin}, to access the particle size distribution).

The most obvious advantage of DDM is that it does not require any dedicated or specialized instrumentation. Other key strengths, which fostered adoption in an increasing number of laboratories worldwide, include its robustness with respect to the presence of dirt along the optical path\cite{cerbino2008differential}, the tolerance with respect to multiple scattering \cite{nixon2022probing}, the possibility of exploiting different imaging modes and optical contrast mechanisms, and even combining them \cite{jepson2013enhanced,giavazzi2018tracking, drechsler2017active} to gather information on different sub-structures.

To date, one of the main limitations of bright-field DDM compared to DLS is its sensitivity, namely the ability to reliably measure very small, weakly scattering particles, a case that includes a variety of interesting biological macromolecules. In this work, we experimentally show that bright-field DDM can successfully probe the dynamics of diluted protein solutions, down to concentrations of a few mg/mL; the smallest detected macromolecule (Lysozyme) has a hydrodynamic radius of about 2 nm, one order of magnitude below the previous detection limit \cite{giavazzi2016structure}, which opens a whole new field of application for this technique.

\section{Materials and methods}

\subsection{Sample preparation and imaging}

In this study, we consider aqueous solutions of three different proteins: Bovine Serum Albumin (BSA)
(Sigma \#A7638), Pepsin (Roche \#10108057001), and Lysozyme (Roche \#10837059001). Phosphate
buffered saline (PBS) was prepared from commercial 10x stock solutions (ROTH \#1058.1). For each protein, we probe four different concentrations, obtained by serial dilution.
Before measurement, each solution is filtered with  0.22 $\mu$m filters (Minisart). 

Filtration and possible protein adsorption to the vials can in principle alter the actual protein concentration in the sample, which can differ from its nominal value.
To account for these effects, a commercial UV spectrophotometer (ThermoFisher) has been used to independently assess the actual protein concentration within each sample. By measuring the sample absorbance $a$ at a wavelength $\lambda=280$ nm, the protein concentration is then obtained by inverting the Beer-Lambert law, $a = \varepsilon c l$, where $\varepsilon$ is the protein extinction coefficient, $c$ is the concentration, and $l=0.1$ cm is the optical path length.
%The most concentrated solutions have been previously diluted before measurement in order to prevent saturation. The measured values are then corrected by the dilution factor to quantify the concentration of the original solution.
The concentrations are reported in Supplementary Table S1, together with the extinction coefficients used for their calculation.
 
% CAPILLARY PREPARATION

For microscopy measurements, the samples are confined in glass capillaries with rectangular cross-section (thickness 300 $\mu$m, width 3 mm).
Each capillary is then put on a microscope slide and carefully sealed with epoxy glue at the two ends to prevent evaporation.

% MICROSCOPY DETAILS

The experimental setup consists of a commercial inverted microscope (Nikon Eclipse Ti-U) equipped with a fast digital camera (Hamamatsu Orca Flash 4.0 v2).
Bright-field images of the sample are collected with a 20x {0.5 NA} objective.
For each sample, four sequences of 5000 images are recorded with a sampling rate of 500 fps and an exposure time of 1.99 ms. For each solution, image sequences are collected from two distinct portions of the sample,  the focal plane corresponding to the middle plane of the capillary.
Image resolution is 1024(w)x128(h) pixels, upon 2x2 binning. The effective pixel size, taking into account lens magnification and binning, is equal to 650 nm. The imaging conditions, in terms of illumination intensity, size of the field diaphragm, and size aperture diaphragm, as well as the camera settings, are carefully maintained constant throughout the whole set of experiments. All the experiments are performed at room temperature $T=(22\pm2)^\circ$C.

%A typical image appears as a noisy background with some large dark spots, corresponding to dust particles along the microscope optical train.
%Indeed, the protein size being much below the diffraction limit, their scattering signal is well below the background and hence its contribution is not visible on the images.

% DLS

All the samples are characterized in parallel  with a state-of-the-art commercial DLS instrument
(Prometheus Panta, NanoTemper Technologies GmbH) with $\lambda = 405$ nm and $\theta = 147^\circ$ for comparison.
Samples were filled into Prometheus High Sensitivity capillaries and 10 acquisitions were measured per capillary at 6\%
UV excitation power and 100\% DLS power at $25.0^\circ$C using PR.Panta Control v.1.5.2.
Additional static light scattering measurements have been performed using a custom-built apparatus equipped with a $\lambda=532$ nm laser source \cite{lago1993quasielastic}.

\subsection{DDM analysis for protein sizing}

\begin{figure*}[htbp]
\centering
\includegraphics[width=\textwidth]{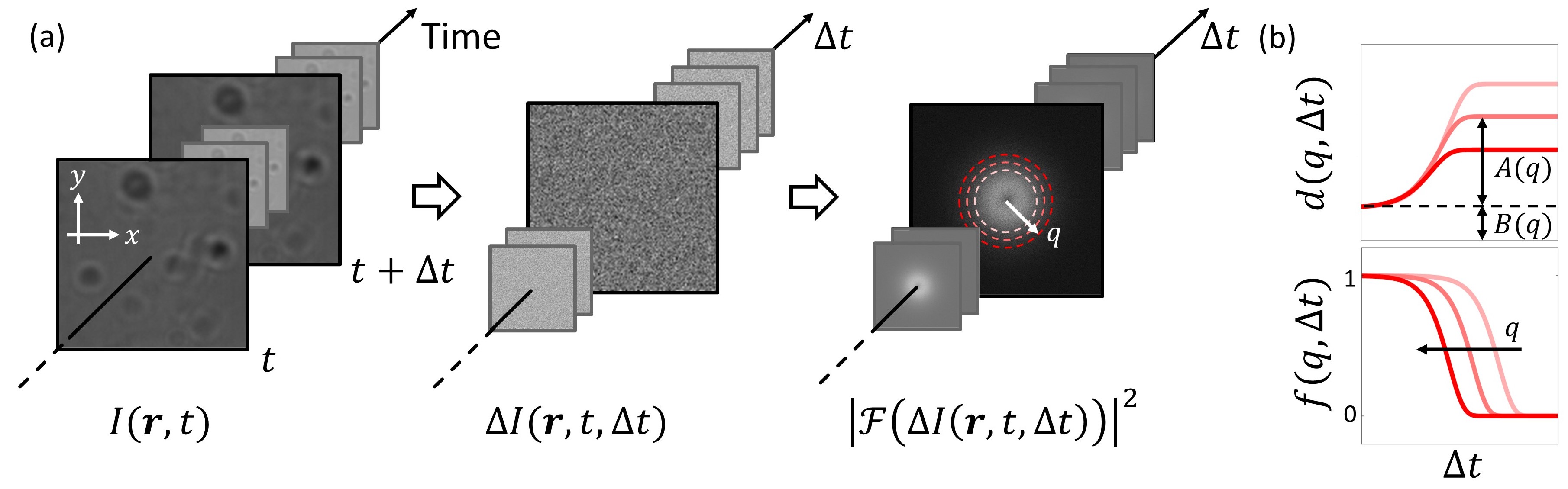}
\caption{\textbf{DDM analysis.} (a) A sequence of sample images $I(\mathbf{r},t)$ is recorded with constant frame rate.
Frames at increasing time delays are subtracted from each reference frame to get difference images $ \Delta I(\mathbf{r},t,\Delta t) = I(\mathbf{r},t+\Delta t) - I(\mathbf{r},t)$.
The Fourier transform of the image difference, and its squared modulus, is then calculated, $|\mathcal{F}({\Delta I}(\mathbf{r},t,\Delta t))|^2$.
The image structure function $d(\mathbf{q},\Delta t)$ is retrieved by averaging the Fourier power spectra obtained for the same time delay $\Delta t$ but different reference times $t$.
The system isotropy allows considering the azimuthal average of $d(\mathbf{q},\Delta t)$.
(b) For each wavevector $q$, the azimuthally-averaged image structure function can be plotted against the lag time $\Delta t$, and can be written as $d(q,\Delta t) = A(q) [ 1 - f(q,\Delta t)] + B(q)$, where $A(q)$ is the static amplitude, $B(q)$ is the noise term, and $f(q,\Delta t)$ is the real part of the intermediate scattering function.}
\label{fig:Figure1_DDM}
\end{figure*}

Protein diffusive dynamics are probed in the reciprocal space according to DDM analysis, whose main steps are schematically illustrated in Figure \ref{fig:Figure1_DDM}. A detailed description of the technique can be found for examples in Refs. \cite{cerbino2008differential,giavazzi2009scattering}, while a number of software implementations of the method have been made publicly available by different groups in the last few years\cite{germain2016differential,verwei2022quantifying}.
In short, the procedure is based on the acquisition of a sequence of bright-field microscopy images $I(\mathbf{r},t)$ with a constant frame rate. Frames separated by a given time delay $\Delta t$ are subtracted from each other to generate a difference image $ \Delta I(\mathbf{r},t,\Delta t) = I(\mathbf{r},t+\Delta t) - I(\mathbf{r},t)$, as depicted in Figure \ref{fig:Figure1_DDM}(a).
A 2D fast Fourier transform algorithm is then applied to the difference image, and the spatial Fourier power spectra obtained for the same time delay $\Delta t$, but different initial times $t$, are then averaged to get the so-called image structure function $ d(\mathbf{q},\Delta t) \equiv \langle | \mathcal{F}({\Delta I}(\mathbf{r},t,\Delta t))|^2\rangle $, where $\mathcal{F}$ indicates the Fourier transform, and $\mathbf{q}=(q_x,q_y)$ is the 2D wavevector in the Fourier space. 
If the system is isotropic, an average can be performed over different orientations of $\mathbf{q}$, obtaining the azimuthally averaged image structure function
$d(q,\Delta t)$, where $q=\sqrt{q_x^2+q_y^2}$.

For each scattering wave-vector $q$, the image structure function $d(q,\Delta t)$ is typically a monotonically increasing function  of the delay time $\Delta t$, as shown in Figure \ref{fig:Figure1_DDM}(b).
%: for  $\Delta t \xrightarrow[]{0}$, $d(q,\Delta t) \xrightarrow[]{}$ increases monotonically, before reaching a plateau at long time delays corresponding to a certai.
The image structure function is given by:
 \begin{equation}
 \label{struct_funct}
  d(q,\Delta t) = A(q) [ 1 - f(q,\Delta t)] + B(q), 
 \end{equation}
 where the term $A(q)$ is the static amplitude, and the term $B(q)$ accounts for the noise in the detection chain \cite{giavazzi2009scattering}; the function $f(q,\Delta t)$ is the real part of the intermediate scattering function, a quantity that is also probed in a DLS experiment for a fixed $q$ set by the scattering angle $\theta=2 \arcsin(q\lambda/\pi)$ \cite{giavazzi2009scattering, berne2000dynamic}.
By fitting a suitable model to $d(q,\Delta t)$, one can thus extract $f(q,\Delta t)$, whose decay encodes the information on the sample dynamics at a lengthscale $2\pi/q$\cite{berne2000dynamic}.

%Thus, performing DDM analysis on the data entails the choice of a suitable model for $f(q,\Delta t)$.
In the simple case where the sample is a collection of monodisperse independent particles undergoing Brownian motion with a diffusion coefficient $D$, the intermediate scattering function is described by a single exponential decay $f(q,\Delta t) =\exp(-\Gamma(q)\Delta t)$, where the $q$-dependent relaxation rate $\Gamma(q)$ is given by $\Gamma(q)=D q^2$ \cite{berne2000dynamic}.
{For each $q$, fitting eq. \ref{struct_funct} to $d(q,\Delta t)$, with $f(q,\Delta t) =\exp(-\Gamma(q)\Delta t)$, allows obtaining the relaxation rate $\Gamma(q)$.}
By fitting a quadratic model to the latter, one estimates the diffusion coefficient, which is linked to the particle hydrodynamic radius $R_h$ via the Stokes-Einstein equation:
\begin{equation}
    D = \frac{k_B T}{6 \pi \eta R_h}
    \label{eq:StokesEinstein}
\end{equation}
where $k_B=1.38 \cdot 10^{-23}$ J/K is the Boltzmann constant, $T$ is the sample absolute temperature, and $\eta$ is the solvent viscosity. This way, one can thus estimate the typical size of the diffusing objects. If the solution contains particles of different sizes, the intermediate scattering function is given by the weighted sum of multiple exponential decays, each of which is associated with a different sub-population of particles \cite{berne2000dynamic}.
In this general case, similarly to DLS, reconstructing the particle size distribution from the intermediate scattering function is a non-trivial task and, according to the complexity of the sample and to the signal-to-noise ratio, different approaches, like cumulant expansion\cite{frisken2001revisiting}
 or inversion algorithms like CONTIN\cite{provencher1982contin}, can be exploited.

\section{Results and Discussion}
\begin{figure}%[htbp]
\centering
\includegraphics[width=.5\columnwidth]{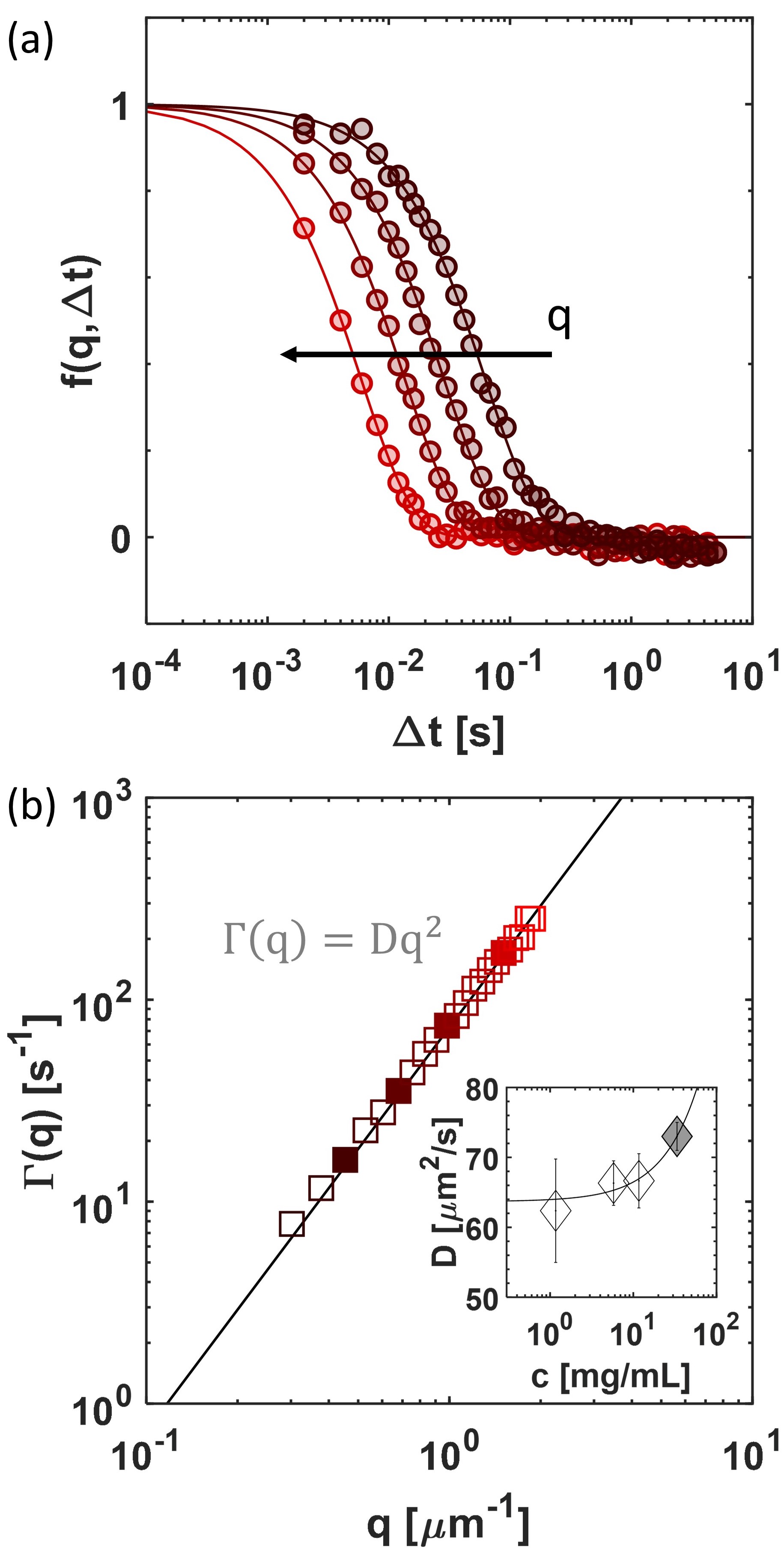}
\caption{\textbf{DDM analysis of diluted BSA solution.} (a) Representative intermediate scattering functions $f(q,\Delta t)$ at different wavevectors $q$ for the BSA sample at concentration $c=34 \pm 2$ mg/mL. The experimental data (circles) are fitted to a single exponential decay (solid lines). (b) Relaxation rate $\Gamma(q)$ extracted from the fit. Solid symbols highlight the wavevectors $q$ corresponding to the curves shown in panel (a). The solid line represents the quadratic fit to the data $\Gamma(q)=Dq^2$. The inset reports the values of the diffusion coefficient $D$ calculated from DDM analysis at different protein concentrations. The value corresponding to the sample at $c=34 \pm 2$ mg/mL is highlighted in gray.
{The solid line represents a linear fit to the data according to eq.\eqref{eq_Dcollective}.}}
\label{fig:Figure2_BSA}
\end{figure}

We performed DDM analysis on all samples, consisting of solutions of three different proteins (Bovine Serum Albumin (BSA), Lysozyme, and Pepsin) at different concentrations, as detailed in the Materials and Methods section.
A comprehensive account of all the collected data can be found in the Supporting Information.  
\textcolor{black}{Representative image sequences are included in Supplementary Movies SM1-SM6. In each movie, both raw and background-subtracted images are shown, while  the corresponding intensity histograms before and after background subtraction are reported in Figures S1 and S2, respectively.}
We consider first the results obtained for the BSA solutions. Figure \ref{fig:Figure2_BSA}(a) shows a few representative intermediate scattering functions $f(q,\Delta t)$, obtained at different wavevectors $q$, for the most concentrated BSA solution ($c=34 \pm2$ mg/mL).
As it can be appreciated from the figure, the experimental data are well described by a single exponential decay.
The $q$-dependence of the corresponding relaxation rate $\Gamma(q)$ obtained from the fit is reported in Figure \ref{fig:Figure2_BSA}(b).
A quadratic fit $\Gamma(q)=D q^2$ provides an estimated diffusion coefficient {$D$=(73 $\pm$ 2) $\mu$m$^{2}$/s.}
%\textcolor{red}{This value is in excellent agreement with the one obtained from DLS on the same sample $D_\text{DLS}$=(68 $\pm$ 5) $\mu$m$^{2}$/s.}
The diffusion coefficients obtained with the same procedure for the BSA samples at lower concentrations (down to about  1.2 mg/mL) are plotted as a function of protein concentration in the inset of Figure \ref{fig:Figure2_BSA}(b).
{The diffusion coefficient is observed to systematically increase with protein concentration. Weak
intermolecular interactions can indeed result in a dependency of the collective diffusion coefficient $D$ on the concentration $c$, which for semi-diluted solutions reads}
{\begin{equation}
D = D_0(1 + k_D \cdot c),
\label{eq_Dcollective}
\end{equation}}
{where $D_0$ is the diffusion coefficient in the dilute limit, and $k_D$ is the so-called diffusion interaction parameter, which is expected to be positive in the case of repulsive interactions \cite{harding1985concentration}.
A linear fit to the data provides $D_0=(64 \pm 4)$ $\mu$m$^2$/s and $k_D = (4 \pm 1) \cdot 10^{-3}$ mL/mg, which is fully consistent with previously reported values for BSA under similar experimental conditions\cite{larsen2021kD}.}

\begin{figure*}[h]
\centering
\includegraphics[width=\textwidth]{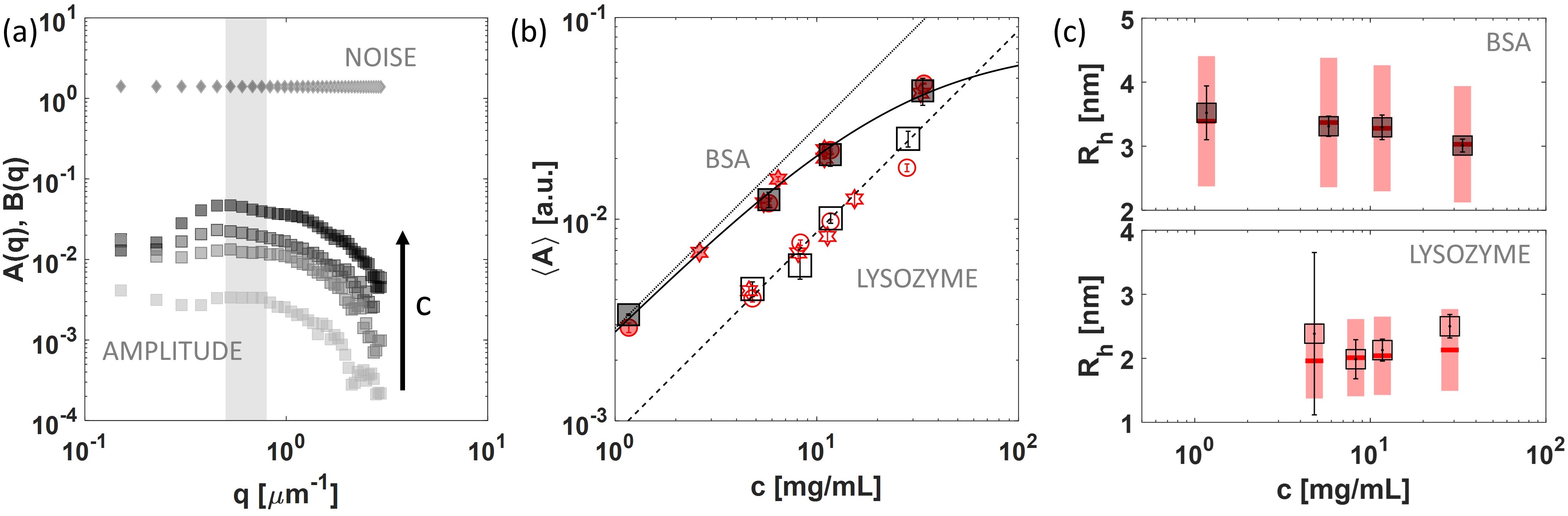}
\caption{\textbf{Signal amplitude and noise at different protein concentrations.} (a) Signal amplitude $A(q)$ (squares) and noise term $B(q)$ (diamonds) measured for BSA solutions at different concentrations.
(b) Scaling of the scattering intensity measured with DDM (black squares) and SLS (red symbols) as a function of protein concentration for BSA (solid symbols) and Lysozyme (empty symbols) samples. DDM values are obtained as an average of $A(q)$ in the $q$ range between 0.5 $\mu$m$^{-1}$ and 0.8 $\mu$m$^{-1}$, corresponding to the gray-shaded area in panel (a). The dashed and dotted lines correspond to linear scalings, while the solid black line is a fit to the BSA DDM data according to eq.\eqref{virial}. Each SLS dataset has been scaled with an arbitrary multiplicative constant. Red circles correspond to SLS measurements at $\lambda=405$ nm and $\theta=147^\circ$, whereas red stars to measurements at $\lambda=532$ nm and $\theta=90^\circ$.
(c) Hydrodynamic radii estimated with DDM (black squares) for single molecules of BSA (top) and Lysozyme (bottom) at different concentrations.
Values obtained from DLS are shown as red horizontal dashes, with vertical shaded bars representing the full width at half maximum of the size distributions.}
\label{fig:Figure3_BSA_LYSO}
\end{figure*}

{To enable a direct comparison with the output of the DLS measurements, we calculated for each sample the corresponding hydrodynamic radius using the Stokes-Einstein relation (eq. \ref{eq:StokesEinstein}). The results are shown in Figure \ref{fig:Figure3_BSA_LYSO}(c).} In the same figure, we also report the average hydrodynamic radius and the width of the number-weighted size distribution estimated for each sample with DLS (see Supplementary Figure S1 and Supplementary Table S2).
As can be appreciated, the agreement between DDM and DLS results, as well as with literature reference values\cite{jachimska2008characterization}, is very good.
{As a further consistency check, we independently estimate the diffusion interaction parameter $k_D$ by fitting the model $R_h^{-1}=R_0^{-1} (1+k_D \cdot c)$ to the DLS data. The fitting procedure gives $R_0 = (3.4 \pm 0.1)$ nm and an estimate of the diffusion interaction parameter $k_D = (4 \pm 1) \cdot 10^{-3}$ mL/mg, fully compatible with the value obtained with DDM.}

Besides extracting the relaxation rate, the above-described procedure also enables estimating the corresponding static amplitude $A(q)$, as well as the noise term $B(q)$, for each wavevector $q$.

In Figure \ref{fig:Figure3_BSA_LYSO}(a) we report the amplitude $A(q)$ and the noise baseline $B(q)$ measured for the BSA solutions at different concentrations.
First of all, we note that the noise term $B(q)$ is nearly constant at all wavevectors and does not differ significantly between the samples. This is consistent with the fact that $B(q)$ mainly depends on the electronic noise of the sensor and on the shot noise \cite{giavazzi2009scattering}, which is expected to be constant as the imaging conditions and the camera settings are identical for all acquisitions.
By contrast, the signal amplitude $A(q)$ systematically increases with protein concentration. Indeed, for each given wavevector $q$, $A(q)$ is expected to be directly proportional to the scattering intensity $I(q)$, $A(q)=T(q)I(q)$, where the proportionality term $T(q)$, often referred to as the optical transfer function, is an instrumental constant \cite{giavazzi2009scattering}, independent of the sample concentration.
To investigate how the amplitude scales with protein concentration, we consider the average value $\langle A\rangle$ over a defined range of $q$ between 0.5 $\mu$m$^{-1}$ and 0.8 $\mu$m$^{-1}$, where $A(q)$ exhibits a very broad peak. 
The results are shown as solid symbols in Figure \ref{fig:Figure3_BSA_LYSO}(b).
As it can be appreciated from the figure, the amplitude displays a fairly linear dependence on concentration for small $c$ (dotted line), while a systematic deviation is observed at larger $c$. This behavior is compatible with the presence of repulsive interactions between the molecules, and it is described by the Debye-Zimm equation\cite{zimm1948scattering} which, for semi-diluted solutions of particles much smaller than the wavelength of light, can be written as
\begin{equation}
    \label{virial_0}
    \frac{K^*c}{\Delta R}=\frac{1}{M}+2B_2 c.
\end{equation}
Here, $\Delta R$ is the Rayleigh ratio, $K^*$ is an instrumental constant (which also incorporates the refractive index of the solvent and the refractive index increment d$n$/d$c$ of the protein solution), $M$ is the molecular mass, and $B_2$ is the second virial coefficient, which represents the first correction to the ideal gas equation of state due to interparticle interactions. 
For a given experimental geometry, the Rayleigh ratio is directly proportional to the scattering intensity $I$. We can thus rewrite eq. \eqref{virial_0} in terms of the average amplitude $\langle A \rangle$ as follows
\begin{equation}
\label{virial}
    \langle A \rangle=\frac{kc}{1+ c/c_0 },
\end{equation}
where $k$ is a new constant accounting also for the effect of the optical transfer function, and $c_0 = (2M B_2)^{-1}$.
By fitting eq. \eqref{virial} to our DDM data (and using $M=66.6$ kDa) we obtain the following estimate for the second virial coefficient {$B_2=(3 \pm 1) \cdot10^{-4}$ mol $\cdot$ mL $\cdot$ g$^{-2}$}.
This value has the same order of magnitude but is significantly larger compared to most of the values reported in the literature for the same protein under similar conditions \cite{ma2015determination}.
In order to confirm the validity of our findings, we performed static light scattering experiments on the same proteins, obtaining the scattered intensity $I(c)$ both at $147^\circ$ ($\lambda=405$ nm) and $90^\circ$ ($\lambda=532$ nm) as a function of the concentration $c$.
The results, scaled by a global multiplicative constant, are reported in Figure \ref{fig:Figure3_BSA_LYSO}(b) as solid red circles and stars respectively, and display  {an excellent} agreement with the ones obtained with DDM.

Taken together, these results demonstrate that bright-field DDM, besides providing an accurate characterization of diluted and semi-diluted protein solution dynamics, can also be used to characterize, at least in relative terms, the dependence of the scattering intensity on the concentration, gathering quantitative information on the intermolecular interactions. 
%An important remark is related to the signal-to-noise ratio. As it can be seen from Figure \ref{fig:Figure3_BSA_LYSO}(a-b), the amplitude of the signal is always far below the noise level, the signal-to-noise ratio ranging from about $1/50$ for the most concentrated sample to about $1/1000$ for the most diluted one.

A similar set of experiments was performed on a sequence of Lysozyme solutions at concentrations ranging between 5 and 30 mg/mL.
%Even in this case,
The obtained intermediate scattering functions are again fully compatible with a simple exponential decay, indicating a fairly monodisperse sample (see Supporting Information). The estimates for the hydrodynamic radius obtained for each sample are reported in Figure \ref{fig:Figure3_BSA_LYSO}(c) as empty symbols,
{where we can see that, within the experimental errors, the estimated $R_h$ values do not show any systematic dependence on concentration. All} values are in good agreement with both the literature \cite{wilkins1999hydrodynamic} and DLS results (see Supplementary Table S2).
For Lysozyme, the average amplitude $\langle A \rangle$ shown in Figure \ref{fig:Figure3_BSA_LYSO}(b) (empty squares) displays a rather clean linear scaling with $c$ (dashed line), indicating a marginal effect of intermolecular interactions in the investigated concentration range. This result was also confirmed by independent static light scattering measurements (empty red circles and stars).

%The results for the and, upon suitable calibration...

%that the amplitude scales in first approximation linearly with the concentration for the Lysozyme samples, while a sublinear dependency is found for BSA solutions. Moreover, larger signal amplitudes are measured for BSA, reflecting the larger molecular size.

%By inverting equation \eqref{eq:StokesEinstein} we can calculate the mean hydrodynamic radius $R_h$ from the measured diffusion coefficient $D$.
%The results are reported in figure \ref{fig:Figure3_BSA_LYSO} for both BSA and Lysozyme.
%The plot shows the excellent agreement between the hydrodynamic radii calculated from DDM analysis and DLS results.

\begin{figure}
\centering
\includegraphics[width=0.8\columnwidth]{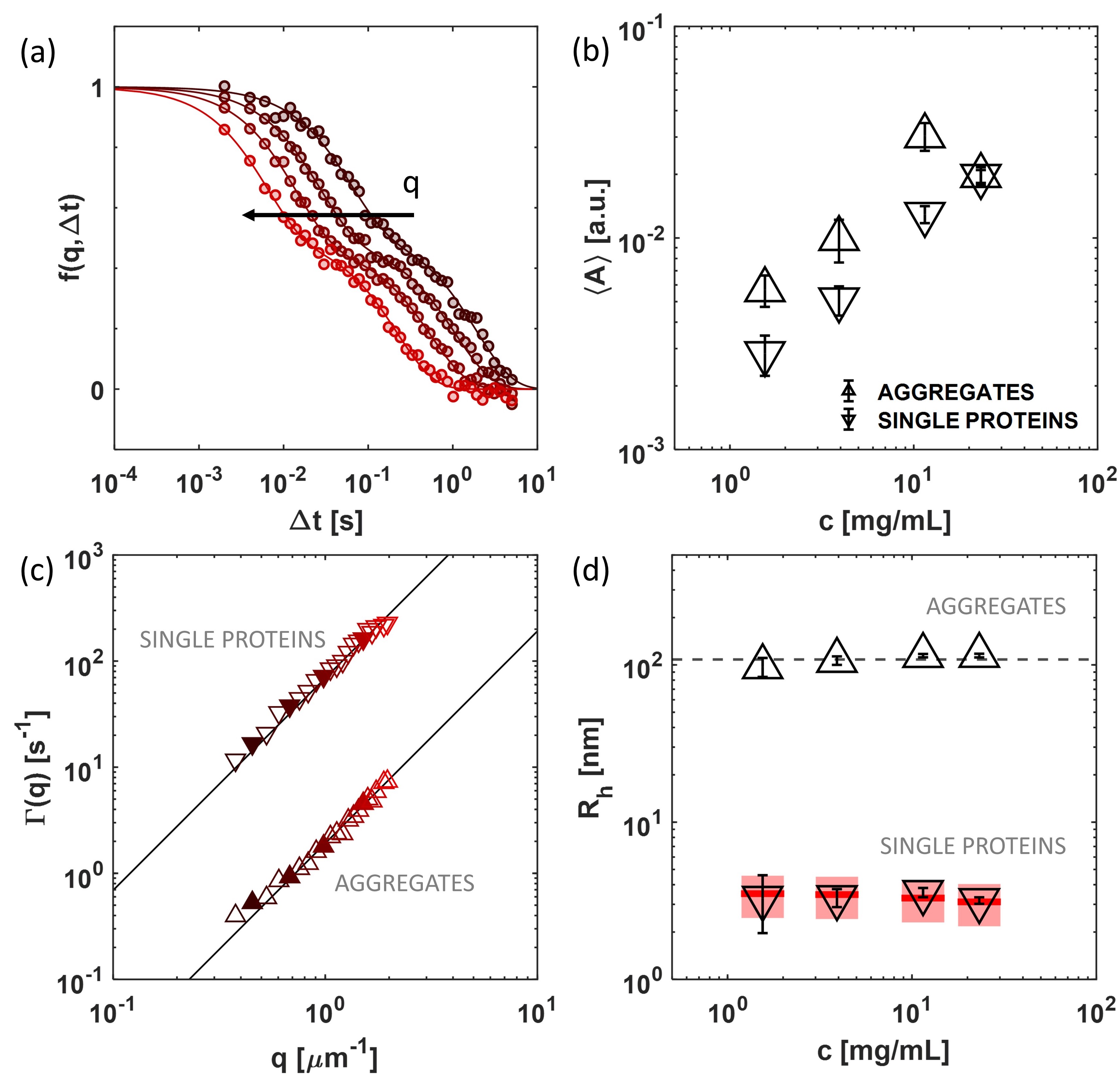}
\caption{\textbf{DDM analysis of diluted Pepsin solutions.} (a) Examples of intermediate scattering functions $f(q,\Delta t)$ at different wavevectors $q$ for the Pepsin sample at $c=23 \pm 1$ mg/mL. Data clearly show a double decay due to the simultaneous presence of single and aggregated proteins. Experimental data (circles) are fitted with a double exponential (solid lines).
(b) Scaling of the two amplitudes $\langle A_1\rangle$ (single proteins) and $\langle A_2 \rangle$ (protein aggregates) with the protein concentration.
(c) Relaxation rates $\Gamma_1$ and $\Gamma_2$ extracted from the fit at different $q$, linked to the motion of single proteins (down-pointing triangles) and protein aggregates (up-pointing triangles) respectively. Solid symbols highlight the selection of data points corresponding to the curves shown in (a). (d) Hydrodynamic radii of single proteins and aggregates, calculated from the diffusion coefficients obtained at different concentrations. The black dashed line represents the average size estimated for the protein aggregates, that is equal to about 108 nm.
The red-shaded bars represent the full width at half maximum of the size distributions obtained from DLS, with average values indicated as red horizontal dashes.}
\label{fig:Figure4_PEP}
\end{figure}

While the results obtained for both BSA and Lysozyme solutions are compatible with the behavior of a diluted suspension of identical particles, the Pepsin solutions we have investigated provide an example of a less ideal case.
Indeed, despite filtration, DDM analysis revealed the presence of protein aggregates at all concentrations considered.
The intermediate scattering functions are no longer described by a single exponential, but display two well-distinct decays, as shown in Figure \ref{fig:Figure4_PEP}(a) for the highest concentration ($c= 23 \pm 1$ mg/mL).
%As it can be appreciated from the figure, 
A good fit to the data is now obtained by adopting a double-exponential model of the kind
$f(q,\Delta t) = \left[1-\alpha(q) e^{-\Gamma_1(q)\Delta t} -\left(1-\alpha(q)\right)e^{-\Gamma_2(q) \Delta t}\right]$.
This model corresponds to the simultaneous presence of two relaxation processes, which we attribute to two distinct families of diffusing particles.
This can be confirmed by considering the $q$-dependence of the two relaxation rates, $\Gamma_1$ and $\Gamma_2$, extracted from the fit, which is shown in Figure \ref{fig:Figure4_PEP}(c).
Both $\Gamma_1$ and $\Gamma_2$ display a rather clean $\sim q^2$ scaling. By fitting a quadratic function to the two rates, we determine the corresponding diffusion coefficients, from which we recover the associated hydrodynamic radii reported in Figure \ref{fig:Figure4_PEP} (d).
Strikingly enough, the hydrodynamic radius $R_{h,1}=(3.2 \pm 0.2)$ nm corresponding to the faster decay is in excellent agreement with the value $R_{h,\text{DLS}}=(3.11 \pm 0.03)$ nm obtained from DLS measurement, as well as with literature values for the monomeric protein under similar conditions \cite{gtari2017impact}.
This indicates that the "fast" contribution to the intermediate scattering function comes from the diffusion of monomeric, unaggregated proteins.
On the other hand, the slower decay provides a hydrodynamic radius $R_{h,2}$ of about $(115 \pm 3)$ nm. This value is compatible with the size of the largest particle that can pass through the filter, whose nominal mesh size is about 220 nm. The amplitude $\langle A_1 \rangle=\langle \alpha \cdot A \rangle$ of the decay corresponding to single proteins and the amplitude $\langle A_2 \rangle= \langle(1-\alpha) \cdot A\rangle$ obtained for the aggregates are both shown in Figure \ref{fig:Figure4_PEP}(b), as down- and up-pointing triangles, respectively. As for Lysozyme, we observe a fairly good linear scaling of $\langle A_1 \rangle$ with $c$ for Pepsin as well over the whole range of investigated concentrations.

It is worth noticing that, while the amplitudes of the two contributions are somehow comparable ($1 \lesssim \langle A_2 \rangle / \langle A_1 \rangle \lesssim 2$, corresponding to $0.3 \lesssim \alpha \lesssim 0.5$), this is not true for the corresponding concentrations $c_1$ and $c_2$. Let $M_1$ and $M_2$ be the average molecular weight of the monomeric proteins and of the aggregates, respectively. Under the simplifying assumptions that both d$n$/d$c$ and the density are the same for single and aggregated molecules, we get
$\frac{c_2}{c_1} \simeq \frac{\langle A_2 \rangle/M_2}{\langle A_1 \rangle/M_1} \simeq \frac{\langle A_2 \rangle}{\langle A_1 \rangle}\left(\frac{R_{h,1}}{R_{h,2}}\right)^3\simeq 8 \cdot 10^{-5}$.

According to this estimate, the contribution of the aggregates to the total mass in the solution is very small (less than 0.01\%) and we can safely assume that the concentration $c_1$ of the monomeric protein coincides with the total concentration $c$ measured with the UV spectrophotometer.
Remarkably, the presence of these aggregates is only barely visible in DLS measurements performed at large scattering angles (see Supplementary Figure S1), which suggests how the access to the small scattering angles probed by DDM may be crucial to obtain a thorough sample characterization.
%The previous discussion illustrates the exquisite sensitivity of DDM to the presence of even a tiny fraction of large particles in a background of smaller ones, similarly to DLS. 

\section{Conclusions}
In this work, we have shown that DDM enables measuring the size of proteins in solution in seconds using a bright-field microscope. The current detection limit in terms of concentration is about 1 mg/mL for BSA, only about a decade above the nominal sensitivity of typical commercial DLS instruments.
The smallest detected molecule is Lysozyme, with a hydrodynamic radius of about $2$ nm. Importantly, these limits are by no means fundamental in nature, as they are mainly determined by the \REV{intrinsic noise of the detector and by the size and number of acquired images.} 
%signal-to-noise ratio of the camera and by the duration of the measurement\cite{giavazzi2014digital}. 
Indeed, we foresee that the adoption of new-generation, better-performing sensors, combined with the execution of more extended measurements could significantly increase the sensitivity of the method, {as well as the accuracy in the determination of the particle size.}
\REV{
%In the section "Estimation of the critical signal-to-noise ratio and of the detection limit"
In Supporting Information, we provide an in-depth examination of the instrumental factors (e.g. sample concentration, molecular weight, and camera specifications) that affect the signal-to-noise ratio for a protein sizing DDM experiment. From this, we derive an equation for the critical signal-to-noise ratio - the threshold below which DDM is unlikely to yield reliable results. This estimation serves a dual purpose: it can be used to predict whether the signal from a particular sample, under the experimental conditions in place, will be strong enough for DDM analysis. Additionally, it can guide the optimization of the experimental protocol, ensuring it is primed for detecting the sample of interest. While these results allow to estimate the lower limit of detection (e.g. in terms of the minimum detectable concentration of a target molecule) an upper limit is imposed by the onset of multiple scattering, which breaks the simple linear relation between the amplitude of the concentration fluctuations in the sample and the optical signal in the images DDM relies on \cite{giavazzi2009scattering}. It's important to note, though, that this constraint mainly applies to extremely turbid samples. Under normal conditions \cite{nixon2022probing}, DDM exhibits the capacity to effectively probe solutions with concentrations up to a thousand times higher than what conventional DLS can manage.
Even within this effective range, it's crucial to keep in mind that interactions between the molecules in solution can affect their collective dynamics and lead to a collective diffusion coefficient $D$ that deviates systematically from the one predicted by the Stokes-Einstein relation Eq. \ref{eq:StokesEinstein}\cite{harding1985concentration,berne2000dynamic}, {as also shown in this study.} As a consequence, DDM, just like DLS, requires suitably diluted solutions to be used as a particle sizing tool. A common empirical approach to determine whether a given solution is diluted enough to enable reliable sizing consists of further diluting the solution (for example by a factor of two), repeating the measurement on the diluted sample, and checking whether the diffusion coefficients obtained from the two samples are mutually consistent \cite{stetefeld2016dynamic}. {Alternatively, the diffusion coefficient in the dilute limit can be obtained by extrapolating to $c=0$ the values obtained at different concentrations, according to eq. (\ref{eq_Dcollective}).}
}
%\textcolor{red}{
%Ideally, the accessible $q$-range is limited from below by the image size $L$ ($q_{\text{min}}=2\pi/L$), and from above by the effective pixel size $a$ ($q_{\text{max}}=\pi/a$ \cite{giavazzi2014digital}.  
%The frame rate $f_0$ of the camera should be fast enough to adequately sample the relaxation of the ISF in the relevant wavevector range, which is typically a subset of the interval $[0.1 1]$ $\mu$m.
%A practical indication could be $f_0 \geq D q_{\text{max}}^2$. Assuming $R_{h}\simeq 2 nm$}

We have also shown that besides determining the relaxation dynamics, with DDM it is also possible to obtain quantitative information on the dependence of the scattered intensity on the concentration and hence on intermolecular interactions, an application that could be pushed further by calibrating the microscope with small colloidal particles to determine the transfer function $T(q)$.

In each experiment presented in this work, the volume probed in a single measurement was as small as $\simeq$ 20 nL. This indicates that our approach, combined with suitably designed sample cells, allows the investigation of extremely small sample volumes. 

Notably, by probing much smaller wavevectors (typically in the range $[0.1, 10]$ $\mu$m$^{-1}$) compared to fixed 90$^\circ$ angle DLS ($q_\text{DLS}\simeq 17$ $\mu$m$^{-1}$, assuming $\lambda= 532$ nm), DDM enables characterizing the dynamics of rapidly diffusing particles with relatively relaxed requirements in terms of sampling frequency. For example, the light scattered at 90$^\circ$ by an aqueous solution of 2 nm particles fluctuates with a characteristic time of about 30 $\mu$s. The same solution probed at $q\simeq 1$ $\mu m^{-1}$ displays fluctuations with a $\simeq 9$ ms lifetime, which can be easily captured with a standard scientific camera \textcolor{red}{}.

In this study, we focused on a few relatively simple representative examples, where the particle size distribution is strongly peaked  around either one or two well-defined values. However, the range of applicability of the method is much wider. As already mentioned, the whole paraphernalia of DLS can be directly exploited to analyze and interpret DDM data: cumulant expansion to evaluate sample polydispersity \cite{frisken2001revisiting,giavazzi2016simultaneous,safari2015differential}, inversion schemes to estimate the particle size distribution, like CONTIN \cite{provencher1982contin}, to mention a few.

%Frase ad effetto, fine.

\section{Acknowledgements}
We acknowledge Thomas Carzaniga for his help with protein concentration measurements \textcolor{black}{and Enrico Lattuada for useful discussions}.
This work was partially supported by the European Space Agency, CORA-MAP TechNES Contract No. 4000128933/19/NL/PG.

\section{Supporting Information}
Supporting Information is available, including:\\
\REV{- estimation of the critical signal-to-noise ratio and of the detection limit}\\
- concentration values measured with UV spectrophotometry;\\
- DLS results;\\
- representative intermediate scattering functions $f(q,\Delta t)$, amplitude $A(q)$ and noise term $B(q)$, and relaxation rate $\Gamma(q)$ for each sample;\\
\REV{- representative movies (SM1-SM6) acquired at the highest and lowest protein concentrations and corresponding intensity histograms.}

\bibliography{bibl}

\providecommand{\latin}[1]{#1}
\makeatletter
\providecommand{\doi}
  {\begingroup\let\do\@makeother\dospecials
  \catcode`\{=1 \catcode`\}=2 \doi@aux}
\providecommand{\doi@aux}[1]{\endgroup\texttt{#1}}
\makeatother
\providecommand*\mcitethebibliography{\thebibliography}
\csname @ifundefined\endcsname{endmcitethebibliography}
  {\let\endmcitethebibliography\endthebibliography}{}
\begin{mcitethebibliography}{43}
\providecommand*\natexlab[1]{#1}
\providecommand*\mciteSetBstSublistMode[1]{}
\providecommand*\mciteSetBstMaxWidthForm[2]{}
\providecommand*\mciteBstWouldAddEndPuncttrue
  {\def\EndOfBibitem{\unskip.}}
\providecommand*\mciteBstWouldAddEndPunctfalse
  {\let\EndOfBibitem\relax}
\providecommand*\mciteSetBstMidEndSepPunct[3]{}
\providecommand*\mciteSetBstSublistLabelBeginEnd[3]{}
\providecommand*\EndOfBibitem{}
\mciteSetBstSublistMode{f}
\mciteSetBstMaxWidthForm{subitem}{(\alph{mcitesubitemcount})}
\mciteSetBstSublistLabelBeginEnd
  {\mcitemaxwidthsubitemform\space}
  {\relax}
  {\relax}

\bibitem[Pecora(1964)]{pecora1964doppler}
Pecora,~R.~d. Doppler shifts in light scattering from pure liquids and polymer
  solutions. \emph{The Journal of Chemical Physics} \textbf{1964}, \emph{40},
  1604--1614\relax
\mciteBstWouldAddEndPuncttrue
\mciteSetBstMidEndSepPunct{\mcitedefaultmidpunct}
{\mcitedefaultendpunct}{\mcitedefaultseppunct}\relax
\EndOfBibitem
\bibitem[Debye(1965)]{debye1965spectral}
Debye,~P. Spectral width of the critical opalescence due to concentration
  fluctuations. \emph{Physical Review Letters} \textbf{1965}, \emph{14},
  783\relax
\mciteBstWouldAddEndPuncttrue
\mciteSetBstMidEndSepPunct{\mcitedefaultmidpunct}
{\mcitedefaultendpunct}{\mcitedefaultseppunct}\relax
\EndOfBibitem
\bibitem[Einstein(1956)]{einstein1956investigations}
Einstein,~A. \emph{Investigations on the Theory of the Brownian Movement};
  Courier Corporation, 1956\relax
\mciteBstWouldAddEndPuncttrue
\mciteSetBstMidEndSepPunct{\mcitedefaultmidpunct}
{\mcitedefaultendpunct}{\mcitedefaultseppunct}\relax
\EndOfBibitem
\bibitem[Dubin \latin{et~al.}(1967)Dubin, Lunacek, and
  Benedek]{dubin1967observation}
Dubin,~S.~B.; Lunacek,~J.~H.; Benedek,~G.~B. Observation of the spectrum of
  light scattered by solutions of biological macromolecules. \emph{Proceedings
  of the National Academy of Sciences} \textbf{1967}, \emph{57},
  1164--1171\relax
\mciteBstWouldAddEndPuncttrue
\mciteSetBstMidEndSepPunct{\mcitedefaultmidpunct}
{\mcitedefaultendpunct}{\mcitedefaultseppunct}\relax
\EndOfBibitem
\bibitem[Foord \latin{et~al.}(1970)Foord, Jakeman, Oliver, Pike, Blagrove,
  Wood, and Peacocke]{foord1970determination}
Foord,~R.; Jakeman,~E.; Oliver,~C.; Pike,~E.; Blagrove,~R.; Wood,~E.;
  Peacocke,~A. Determination of diffusion coefficients of haemocyanin at low
  concentration by intensity fluctuation spectroscopy of scattered laser light.
  \emph{Nature} \textbf{1970}, \emph{227}, 242--245\relax
\mciteBstWouldAddEndPuncttrue
\mciteSetBstMidEndSepPunct{\mcitedefaultmidpunct}
{\mcitedefaultendpunct}{\mcitedefaultseppunct}\relax
\EndOfBibitem
\bibitem[Stetefeld \latin{et~al.}(2016)Stetefeld, McKenna, and
  Patel]{stetefeld2016dynamic}
Stetefeld,~J.; McKenna,~S.~A.; Patel,~T.~R. Dynamic light scattering: a
  practical guide and applications in biomedical sciences. \emph{Biophysical
  reviews} \textbf{2016}, \emph{8}, 409--427\relax
\mciteBstWouldAddEndPuncttrue
\mciteSetBstMidEndSepPunct{\mcitedefaultmidpunct}
{\mcitedefaultendpunct}{\mcitedefaultseppunct}\relax
\EndOfBibitem
\bibitem[Hulst and van~de Hulst(1981)Hulst, and van~de Hulst]{hulst1981light}
Hulst,~H.~C.; van~de Hulst,~H.~C. \emph{Light scattering by small particles};
  Courier Corporation, 1981\relax
\mciteBstWouldAddEndPuncttrue
\mciteSetBstMidEndSepPunct{\mcitedefaultmidpunct}
{\mcitedefaultendpunct}{\mcitedefaultseppunct}\relax
\EndOfBibitem
\bibitem[Pusey(1999)]{pusey1999suppression}
Pusey,~P. Suppression of multiple scattering by photon cross-correlation
  techniques. \emph{Current opinion in colloid \& interface science}
  \textbf{1999}, \emph{4}, 177--185\relax
\mciteBstWouldAddEndPuncttrue
\mciteSetBstMidEndSepPunct{\mcitedefaultmidpunct}
{\mcitedefaultendpunct}{\mcitedefaultseppunct}\relax
\EndOfBibitem
\bibitem[Pine \latin{et~al.}(1988)Pine, Weitz, Chaikin, and
  Herbolzheimer]{PhysRevLett.60.1134}
Pine,~D.~J.; Weitz,~D.~A.; Chaikin,~P.~M.; Herbolzheimer,~E. Diffusing wave
  spectroscopy. \emph{Phys. Rev. Lett.} \textbf{1988}, \emph{60},
  1134--1137\relax
\mciteBstWouldAddEndPuncttrue
\mciteSetBstMidEndSepPunct{\mcitedefaultmidpunct}
{\mcitedefaultendpunct}{\mcitedefaultseppunct}\relax
\EndOfBibitem
\bibitem[Cerbino and Trappe(2008)Cerbino, and Trappe]{cerbino2008differential}
Cerbino,~R.; Trappe,~V. Differential dynamic microscopy: probing wave vector
  dependent dynamics with a microscope. \emph{Physical review letters}
  \textbf{2008}, \emph{100}, 188102\relax
\mciteBstWouldAddEndPuncttrue
\mciteSetBstMidEndSepPunct{\mcitedefaultmidpunct}
{\mcitedefaultendpunct}{\mcitedefaultseppunct}\relax
\EndOfBibitem
\bibitem[Giavazzi \latin{et~al.}(2009)Giavazzi, Brogioli, Trappe, Bellini, and
  Cerbino]{giavazzi2009scattering}
Giavazzi,~F.; Brogioli,~D.; Trappe,~V.; Bellini,~T.; Cerbino,~R. Scattering
  information obtained by optical microscopy: differential dynamic microscopy
  and beyond. \emph{Physical Review E} \textbf{2009}, \emph{80}, 031403\relax
\mciteBstWouldAddEndPuncttrue
\mciteSetBstMidEndSepPunct{\mcitedefaultmidpunct}
{\mcitedefaultendpunct}{\mcitedefaultseppunct}\relax
\EndOfBibitem
\bibitem[Bayles \latin{et~al.}(2016)Bayles, Squires, and
  Helgeson]{bayles2016dark}
Bayles,~A.~V.; Squires,~T.~M.; Helgeson,~M.~E. Dark-field differential dynamic
  microscopy. \emph{Soft Matter} \textbf{2016}, \emph{12}, 2440--2452\relax
\mciteBstWouldAddEndPuncttrue
\mciteSetBstMidEndSepPunct{\mcitedefaultmidpunct}
{\mcitedefaultendpunct}{\mcitedefaultseppunct}\relax
\EndOfBibitem
\bibitem[Lu \latin{et~al.}(2012)Lu, Giavazzi, Angelini, Zaccarelli, Jargstorff,
  Schofield, Wilking, Romanowsky, Weitz, and Cerbino]{lu2012characterizing}
Lu,~P.~J.; Giavazzi,~F.; Angelini,~T.~E.; Zaccarelli,~E.; Jargstorff,~F.;
  Schofield,~A.~B.; Wilking,~J.~N.; Romanowsky,~M.~B.; Weitz,~D.~A.;
  Cerbino,~R. Characterizing concentrated, multiply scattering, and actively
  driven fluorescent systems with confocal differential dynamic microscopy.
  \emph{Physical review letters} \textbf{2012}, \emph{108}, 218103\relax
\mciteBstWouldAddEndPuncttrue
\mciteSetBstMidEndSepPunct{\mcitedefaultmidpunct}
{\mcitedefaultendpunct}{\mcitedefaultseppunct}\relax
\EndOfBibitem
\bibitem[Brizioli \latin{et~al.}(2022)Brizioli, Sentjabrskaja, Egelhaaf,
  Laurati, Cerbino, and Giavazzi]{brizioli2022reciprocal}
Brizioli,~M.; Sentjabrskaja,~T.; Egelhaaf,~S.~U.; Laurati,~M.; Cerbino,~R.;
  Giavazzi,~F. Reciprocal space study of Brownian yet non-Gaussian diffusion of
  small tracers in a hard-sphere glass. \emph{Frontiers in Physics}
  \textbf{2022}, 408\relax
\mciteBstWouldAddEndPuncttrue
\mciteSetBstMidEndSepPunct{\mcitedefaultmidpunct}
{\mcitedefaultendpunct}{\mcitedefaultseppunct}\relax
\EndOfBibitem
\bibitem[Bayles \latin{et~al.}(2017)Bayles, Squires, and
  Helgeson]{bayles2017probe}
Bayles,~A.~V.; Squires,~T.~M.; Helgeson,~M.~E. Probe microrheology without
  particle tracking by differential dynamic microscopy. \emph{Rheologica Acta}
  \textbf{2017}, \emph{56}, 863--869\relax
\mciteBstWouldAddEndPuncttrue
\mciteSetBstMidEndSepPunct{\mcitedefaultmidpunct}
{\mcitedefaultendpunct}{\mcitedefaultseppunct}\relax
\EndOfBibitem
\bibitem[Edera \latin{et~al.}(2017)Edera, Bergamini, Trappe, Giavazzi, and
  Cerbino]{edera2017differential}
Edera,~P.; Bergamini,~D.; Trappe,~V.; Giavazzi,~F.; Cerbino,~R. Differential
  dynamic microscopy microrheology of soft materials: A tracking-free
  determination of the frequency-dependent loss and storage moduli.
  \emph{Physical Review Materials} \textbf{2017}, \emph{1}, 073804\relax
\mciteBstWouldAddEndPuncttrue
\mciteSetBstMidEndSepPunct{\mcitedefaultmidpunct}
{\mcitedefaultendpunct}{\mcitedefaultseppunct}\relax
\EndOfBibitem
\bibitem[Cerbino \latin{et~al.}(2022)Cerbino, Giavazzi, and
  Helgeson]{cerbino2022differential}
Cerbino,~R.; Giavazzi,~F.; Helgeson,~M.~E. Differential dynamic microscopy for
  the characterization of polymer systems. \emph{Journal of Polymer Science}
  \textbf{2022}, \emph{60}, 1079--1089\relax
\mciteBstWouldAddEndPuncttrue
\mciteSetBstMidEndSepPunct{\mcitedefaultmidpunct}
{\mcitedefaultendpunct}{\mcitedefaultseppunct}\relax
\EndOfBibitem
\bibitem[Wilson \latin{et~al.}(2011)Wilson, Martinez, Schwarz-Linek, Tailleur,
  Bryant, Pusey, and Poon]{wilson2011differential}
Wilson,~L.~G.; Martinez,~V.~A.; Schwarz-Linek,~J.; Tailleur,~J.; Bryant,~G.;
  Pusey,~P.; Poon,~W.~C. Differential dynamic microscopy of bacterial motility.
  \emph{Physical review letters} \textbf{2011}, \emph{106}, 018101\relax
\mciteBstWouldAddEndPuncttrue
\mciteSetBstMidEndSepPunct{\mcitedefaultmidpunct}
{\mcitedefaultendpunct}{\mcitedefaultseppunct}\relax
\EndOfBibitem
\bibitem[Martinez \latin{et~al.}(2012)Martinez, Besseling, Croze, Tailleur,
  Reufer, Schwarz-Linek, Wilson, Bees, and Poon]{martinez2012differential}
Martinez,~V.~A.; Besseling,~R.; Croze,~O.~A.; Tailleur,~J.; Reufer,~M.;
  Schwarz-Linek,~J.; Wilson,~L.~G.; Bees,~M.~A.; Poon,~W.~C. Differential
  dynamic microscopy: A high-throughput method for characterizing the motility
  of microorganisms. \emph{Biophysical journal} \textbf{2012}, \emph{103},
  1637--1647\relax
\mciteBstWouldAddEndPuncttrue
\mciteSetBstMidEndSepPunct{\mcitedefaultmidpunct}
{\mcitedefaultendpunct}{\mcitedefaultseppunct}\relax
\EndOfBibitem
\bibitem[Giavazzi \latin{et~al.}(2018)Giavazzi, Malinverno, Scita, and
  Cerbino]{giavazzi2018tracking}
Giavazzi,~F.; Malinverno,~C.; Scita,~G.; Cerbino,~R. Tracking-free
  determination of single-cell displacements and division rates in confluent
  monolayers. \emph{Frontiers in Physics} \textbf{2018}, \emph{6}, 120\relax
\mciteBstWouldAddEndPuncttrue
\mciteSetBstMidEndSepPunct{\mcitedefaultmidpunct}
{\mcitedefaultendpunct}{\mcitedefaultseppunct}\relax
\EndOfBibitem
\bibitem[Latreille \latin{et~al.}(2022)Latreille, Rabanel, Le~Goas, Salimi,
  Arlt, Patten, Ramassamy, Hildgen, Martinez, and Banquy]{latreille2022situ}
Latreille,~P.-L.; Rabanel,~J.-M.; Le~Goas,~M.; Salimi,~S.; Arlt,~J.;
  Patten,~S.~A.; Ramassamy,~C.; Hildgen,~P.; Martinez,~V.~A.; Banquy,~X. In
  Situ Characterization of the Protein Corona of Nanoparticles In Vitro and In
  Vivo. \emph{Advanced Materials} \textbf{2022}, 2203354\relax
\mciteBstWouldAddEndPuncttrue
\mciteSetBstMidEndSepPunct{\mcitedefaultmidpunct}
{\mcitedefaultendpunct}{\mcitedefaultseppunct}\relax
\EndOfBibitem
\bibitem[Safari \latin{et~al.}(2015)Safari, Vorontsova, Poling-Skutvik,
  Vekilov, and Conrad]{safari2015differential}
Safari,~M.~S.; Vorontsova,~M.~A.; Poling-Skutvik,~R.; Vekilov,~P.~G.;
  Conrad,~J.~C. Differential dynamic microscopy of weakly scattering and
  polydisperse protein-rich clusters. \emph{Physical Review E} \textbf{2015},
  \emph{92}, 042712\relax
\mciteBstWouldAddEndPuncttrue
\mciteSetBstMidEndSepPunct{\mcitedefaultmidpunct}
{\mcitedefaultendpunct}{\mcitedefaultseppunct}\relax
\EndOfBibitem
\bibitem[Giavazzi \latin{et~al.}(2014)Giavazzi, Crotti, Speciale, Serra,
  Zanchetta, Trappe, Buscaglia, Bellini, and
  Cerbino]{giavazzi2014viscoelasticity}
Giavazzi,~F.; Crotti,~S.; Speciale,~A.; Serra,~F.; Zanchetta,~G.; Trappe,~V.;
  Buscaglia,~M.; Bellini,~T.; Cerbino,~R. Viscoelasticity of nematic liquid
  crystals at a glance. \emph{Soft Matter} \textbf{2014}, \emph{10},
  3938--3949\relax
\mciteBstWouldAddEndPuncttrue
\mciteSetBstMidEndSepPunct{\mcitedefaultmidpunct}
{\mcitedefaultendpunct}{\mcitedefaultseppunct}\relax
\EndOfBibitem
\bibitem[Giavazzi \latin{et~al.}(2016)Giavazzi, Haro-P{\'e}rez, and
  Cerbino]{giavazzi2016simultaneous}
Giavazzi,~F.; Haro-P{\'e}rez,~C.; Cerbino,~R. Simultaneous characterization of
  rotational and translational diffusion of optically anisotropic particles by
  optical microscopy. \emph{Journal of Physics: Condensed Matter}
  \textbf{2016}, \emph{28}, 195201\relax
\mciteBstWouldAddEndPuncttrue
\mciteSetBstMidEndSepPunct{\mcitedefaultmidpunct}
{\mcitedefaultendpunct}{\mcitedefaultseppunct}\relax
\EndOfBibitem
\bibitem[Cerbino \latin{et~al.}(2017)Cerbino, Piotti, Buscaglia, and
  Giavazzi]{cerbino2017dark}
Cerbino,~R.; Piotti,~D.; Buscaglia,~M.; Giavazzi,~F. Dark field differential
  dynamic microscopy enables accurate characterization of the
  roto-translational dynamics of bacteria and colloidal clusters. \emph{Journal
  of Physics: Condensed Matter} \textbf{2017}, \emph{30}, 025901\relax
\mciteBstWouldAddEndPuncttrue
\mciteSetBstMidEndSepPunct{\mcitedefaultmidpunct}
{\mcitedefaultendpunct}{\mcitedefaultseppunct}\relax
\EndOfBibitem
\bibitem[Provencher(1982)]{provencher1982contin}
Provencher,~S.~W. CONTIN: a general purpose constrained regularization program
  for inverting noisy linear algebraic and integral equations. \emph{Computer
  Physics Communications} \textbf{1982}, \emph{27}, 229--242\relax
\mciteBstWouldAddEndPuncttrue
\mciteSetBstMidEndSepPunct{\mcitedefaultmidpunct}
{\mcitedefaultendpunct}{\mcitedefaultseppunct}\relax
\EndOfBibitem
\bibitem[Nixon-Luke \latin{et~al.}(2022)Nixon-Luke, Arlt, Poon, Bryant, and
  Martinez]{nixon2022probing}
Nixon-Luke,~R.; Arlt,~J.; Poon,~W.~C.; Bryant,~G.; Martinez,~V.~A. Probing the
  dynamics of turbid colloidal suspensions using differential dynamic
  microscopy. \emph{Soft Matter} \textbf{2022}, \emph{18}, 1858--1867\relax
\mciteBstWouldAddEndPuncttrue
\mciteSetBstMidEndSepPunct{\mcitedefaultmidpunct}
{\mcitedefaultendpunct}{\mcitedefaultseppunct}\relax
\EndOfBibitem
\bibitem[Jepson \latin{et~al.}(2013)Jepson, Martinez, Schwarz-Linek, Morozov,
  and Poon]{jepson2013enhanced}
Jepson,~A.; Martinez,~V.~A.; Schwarz-Linek,~J.; Morozov,~A.; Poon,~W.~C.
  Enhanced diffusion of nonswimmers in a three-dimensional bath of motile
  bacteria. \emph{Physical Review E} \textbf{2013}, \emph{88}, 041002\relax
\mciteBstWouldAddEndPuncttrue
\mciteSetBstMidEndSepPunct{\mcitedefaultmidpunct}
{\mcitedefaultendpunct}{\mcitedefaultseppunct}\relax
\EndOfBibitem
\bibitem[Drechsler \latin{et~al.}(2017)Drechsler, Giavazzi, Cerbino, and
  Palacios]{drechsler2017active}
Drechsler,~M.; Giavazzi,~F.; Cerbino,~R.; Palacios,~I.~M. Active diffusion and
  advection in Drosophila oocytes result from the interplay of actin and
  microtubules. \emph{Nature communications} \textbf{2017}, \emph{8},
  1520\relax
\mciteBstWouldAddEndPuncttrue
\mciteSetBstMidEndSepPunct{\mcitedefaultmidpunct}
{\mcitedefaultendpunct}{\mcitedefaultseppunct}\relax
\EndOfBibitem
\bibitem[Giavazzi \latin{et~al.}(2016)Giavazzi, Savorana, Vailati, and
  Cerbino]{giavazzi2016structure}
Giavazzi,~F.; Savorana,~G.; Vailati,~A.; Cerbino,~R. Structure and dynamics of
  concentration fluctuations in a non-equilibrium dense colloidal suspension.
  \emph{Soft Matter} \textbf{2016}, \emph{12}, 6588--6600\relax
\mciteBstWouldAddEndPuncttrue
\mciteSetBstMidEndSepPunct{\mcitedefaultmidpunct}
{\mcitedefaultendpunct}{\mcitedefaultseppunct}\relax
\EndOfBibitem
\bibitem[Lago \latin{et~al.}(1993)Lago, Rovati, Cant{\`u}, and
  Corti]{lago1993quasielastic}
Lago,~P.; Rovati,~L.; Cant{\`u},~L.; Corti,~M. A quasielastic light scattering
  detector for chromatographic analysis. \emph{Review of scientific
  instruments} \textbf{1993}, \emph{64}, 1797--1802\relax
\mciteBstWouldAddEndPuncttrue
\mciteSetBstMidEndSepPunct{\mcitedefaultmidpunct}
{\mcitedefaultendpunct}{\mcitedefaultseppunct}\relax
\EndOfBibitem
\bibitem[Germain \latin{et~al.}(2016)Germain, Leocmach, and
  Gibaud]{germain2016differential}
Germain,~D.; Leocmach,~M.; Gibaud,~T. Differential dynamic microscopy to
  characterize Brownian motion and bacteria motility. \emph{American Journal of
  Physics} \textbf{2016}, \emph{84}, 202--210\relax
\mciteBstWouldAddEndPuncttrue
\mciteSetBstMidEndSepPunct{\mcitedefaultmidpunct}
{\mcitedefaultendpunct}{\mcitedefaultseppunct}\relax
\EndOfBibitem
\bibitem[Verwei \latin{et~al.}(2022)Verwei, Lee, Leech, Petitjean, Koenderink,
  Robertson-Anderson, and McGorty]{verwei2022quantifying}
Verwei,~H.~N.; Lee,~G.; Leech,~G.; Petitjean,~I.~I.; Koenderink,~G.~H.;
  Robertson-Anderson,~R.~M.; McGorty,~R.~J. Quantifying cytoskeleton dynamics
  using differential dynamic microscopy. \emph{JoVE (Journal of Visualized
  Experiments)} \textbf{2022}, e63931\relax
\mciteBstWouldAddEndPuncttrue
\mciteSetBstMidEndSepPunct{\mcitedefaultmidpunct}
{\mcitedefaultendpunct}{\mcitedefaultseppunct}\relax
\EndOfBibitem
\bibitem[Berne and Pecora(2000)Berne, and Pecora]{berne2000dynamic}
Berne,~B.~J.; Pecora,~R. \emph{Dynamic light scattering: with applications to
  chemistry, biology, and physics}; Courier Corporation, 2000\relax
\mciteBstWouldAddEndPuncttrue
\mciteSetBstMidEndSepPunct{\mcitedefaultmidpunct}
{\mcitedefaultendpunct}{\mcitedefaultseppunct}\relax
\EndOfBibitem
\bibitem[Frisken(2001)]{frisken2001revisiting}
Frisken,~B.~J. Revisiting the method of cumulants for the analysis of dynamic
  light-scattering data. \emph{Applied optics} \textbf{2001}, \emph{40},
  4087--4091\relax
\mciteBstWouldAddEndPuncttrue
\mciteSetBstMidEndSepPunct{\mcitedefaultmidpunct}
{\mcitedefaultendpunct}{\mcitedefaultseppunct}\relax
\EndOfBibitem
\bibitem[Harding and Johnson(1985)Harding, and
  Johnson]{harding1985concentration}
Harding,~S.~E.; Johnson,~P. The concentration-dependence of macromolecular
  parameters. \emph{Biochemical Journal} \textbf{1985}, \emph{231},
  543--547\relax
\mciteBstWouldAddEndPuncttrue
\mciteSetBstMidEndSepPunct{\mcitedefaultmidpunct}
{\mcitedefaultendpunct}{\mcitedefaultseppunct}\relax
\EndOfBibitem
\bibitem[Larsen \latin{et~al.}(2021)Larsen, Atkins, and Nath]{larsen2021kD}
Larsen,~H.~A.; Atkins,~W.~M.; Nath,~A. Probing interactions of therapeutic
  antibodies with serum via second virial coefficient measurements.
  \emph{Biophysical Journal} \textbf{2021}, \emph{120}, 4067--4078\relax
\mciteBstWouldAddEndPuncttrue
\mciteSetBstMidEndSepPunct{\mcitedefaultmidpunct}
{\mcitedefaultendpunct}{\mcitedefaultseppunct}\relax
\EndOfBibitem
\bibitem[Jachimska \latin{et~al.}(2008)Jachimska, Wasilewska, and
  Adamczyk]{jachimska2008characterization}
Jachimska,~B.; Wasilewska,~M.; Adamczyk,~Z. Characterization of globular
  protein solutions by dynamic light scattering, electrophoretic mobility, and
  viscosity measurements. \emph{Langmuir} \textbf{2008}, \emph{24},
  6866--6872\relax
\mciteBstWouldAddEndPuncttrue
\mciteSetBstMidEndSepPunct{\mcitedefaultmidpunct}
{\mcitedefaultendpunct}{\mcitedefaultseppunct}\relax
\EndOfBibitem
\bibitem[Zimm(1948)]{zimm1948scattering}
Zimm,~B.~H. The scattering of light and the radial distribution function of
  high polymer solutions. \emph{The Journal of chemical physics} \textbf{1948},
  \emph{16}, 1093--1099\relax
\mciteBstWouldAddEndPuncttrue
\mciteSetBstMidEndSepPunct{\mcitedefaultmidpunct}
{\mcitedefaultendpunct}{\mcitedefaultseppunct}\relax
\EndOfBibitem
\bibitem[Ma \latin{et~al.}(2015)Ma, Acosta, Whitney, Podgornik, Steinmetz,
  French, and Parsegian]{ma2015determination}
Ma,~Y.; Acosta,~D.~M.; Whitney,~J.~R.; Podgornik,~R.; Steinmetz,~N.~F.;
  French,~R.~H.; Parsegian,~V.~A. Determination of the second virial
  coefficient of bovine serum albumin under varying pH and ionic strength by
  composition-gradient multi-angle static light scattering. \emph{Journal of
  biological physics} \textbf{2015}, \emph{41}, 85--97\relax
\mciteBstWouldAddEndPuncttrue
\mciteSetBstMidEndSepPunct{\mcitedefaultmidpunct}
{\mcitedefaultendpunct}{\mcitedefaultseppunct}\relax
\EndOfBibitem
\bibitem[Wilkins \latin{et~al.}(1999)Wilkins, Grimshaw, Receveur, Dobson,
  Jones, and Smith]{wilkins1999hydrodynamic}
Wilkins,~D.~K.; Grimshaw,~S.~B.; Receveur,~V.; Dobson,~C.~M.; Jones,~J.~A.;
  Smith,~L.~J. Hydrodynamic radii of native and denatured proteins measured by
  pulse field gradient NMR techniques. \emph{Biochemistry} \textbf{1999},
  \emph{38}, 16424--16431\relax
\mciteBstWouldAddEndPuncttrue
\mciteSetBstMidEndSepPunct{\mcitedefaultmidpunct}
{\mcitedefaultendpunct}{\mcitedefaultseppunct}\relax
\EndOfBibitem
\bibitem[Gtari \latin{et~al.}(2017)Gtari, Bey, Aschi, Bitri, and
  Othman]{gtari2017impact}
Gtari,~W.; Bey,~H.; Aschi,~A.; Bitri,~L.; Othman,~T. Impact of macromolecular
  crowding on structure and properties of pepsin and trypsin. \emph{Materials
  Science and Engineering: C} \textbf{2017}, \emph{72}, 98--105\relax
\mciteBstWouldAddEndPuncttrue
\mciteSetBstMidEndSepPunct{\mcitedefaultmidpunct}
{\mcitedefaultendpunct}{\mcitedefaultseppunct}\relax
\EndOfBibitem
\end{mcitethebibliography}

\end{document}

\maketitle

\renewcommand\thefigure{S\arabic{figure}}
\setcounter{figure}{0}

\renewcommand\thetable{S\arabic{table}}
\setcounter{table}{0}

\renewcommand{\theequation}{S\arabic{equation}}

\REV{\section{Estimation of the critical signal-to-noise ratio and of the detection limit}}

\REV{
In this section, we derive a simplified expression to calculate the critical signal-to-noise ratio, below which our method may cease to provide reliable sample dynamics estimates.
Here, the signal-to-noise ratio (SNR) is defined as the ratio between the mean square amplitude $A(q)$ (intensity fluctuation linked to the diffusing molecules at a particular wavevector $q$) and the mean square amplitude $B$ of the noise.
In the final part of the section, we also provide practical guidance on estimating the SNR value for a particular combination of experimental parameters. The equations derived in this section serve a dual purpose.
Firstly, they can predict whether DDM will be suitable for characterizing a specific sample under given experimental conditions.
Secondly, they provide recommendations for optimizing the experimental protocol to achieve the necessary performance for detecting a sample of interest.
We start by representing the intensity distribution in each image $I(\mathbf{x},t)$ as the sum of three components,
\begin{equation*} I(\mathbf{x},t)=I_0(\mathbf{x})+S(\mathbf{x},t)+N(\mathbf{x},t)  ,
\end{equation*}
where $I_0(\mathbf{x})$ is a time-independent background, $S(\mathbf{x},t)$ represents the intensity fluctuation associated with diffusing molecules, and $N(\mathbf{x},t)$ is the detection noise.
The Fourier transform of the difference $\Delta I(\mathbf{x},t,\Delta t)$ between two images taken at times $t+\Delta t$ and $t$ respectively, can be written as follows
\begin{equation*}
\Delta \hat{I}(\mathbf{q},t,\Delta t)=\Delta \hat{S}(\mathbf{q},t,\Delta t)+\Delta \hat{N}(\mathbf{q},t,\Delta t).
\end{equation*}
We assume that $\Delta \hat{S}(\mathbf{q},t,\Delta t)$ and $\Delta \hat{N}(\mathbf{q},t,\Delta t)$ are independent complex Gaussian variables with zero mean.
By introducing the image structure function $d(q,t)=\langle |\Delta \hat{I}(\mathbf{q},t,\Delta t)|^2 \rangle $ as per the definition given in the main text, we obtain
\begin{equation*}
d(q,t)=\langle{|\Delta \hat{S}(\mathbf{q},t,\Delta t)|^2}\rangle+
\langle{|\Delta \hat{N}(\mathbf{q},t,\Delta t)|^2}\rangle,
\end{equation*}
where (as per Eq.1 in the main text)
\begin{equation*}
\langle{|\Delta \hat{S}(\mathbf{q},t,\Delta t)|^2}\rangle=A(q) [1-f(q,\Delta t)],
\end{equation*}
and
\begin{equation*}
\langle{|\Delta \hat{N}(\mathbf{q},t,\Delta t)|^2}\rangle=B.
\end{equation*}
We can also calculate $d_2(q,t)=\langle |\Delta \hat{I}(\mathbf{q},t,\Delta t)|^4 \rangle $
\begin{equation*}
 d_2(q,\Delta t)=\langle{|\Delta \hat{S}(\mathbf{q},t,\Delta t)|^4}\rangle+
 \langle{|\Delta \hat{N}(\mathbf{q},t,\Delta t)|^4}\rangle+4\langle{|\Delta \hat{S}(\mathbf{q},t,\Delta t)|^2}\rangle \langle{|\Delta \hat{N}(\mathbf{q},t,\Delta t)|^2}\rangle.
 \end{equation*}
Since the relation $\langle |g|^4 \rangle =2 \langle |g|^2 \rangle^2 $ holds for a generic complex Gaussian variable $g$, the last equation can be rewritten as
\begin{equation*}
 d_2(q,\Delta t)=2A^2(q) [1-f(q,\Delta t)]^2 + 2B^2 +4 A(q) B [1-f(q,\Delta t)]\simeq 2 B^2,
 \end{equation*}
where the last equivalence holds if $A(q) \ll B$, a condition that is always fulfilled  for the measurements presented in this work.
In experiments, the image structure function is estimated as the mean value of $|\Delta \hat{I}(\mathbf{q},t,\Delta t)|^2$ over a finite number $N_{\Delta t}$ of image differences and a finite number $N_q$ of 2-dimensional wavevectors $\mathbf{q}$ corresponding to the same modulus $q$. For example, if the azimuthal average in the Fourier plane is performed by considering ring-shaped regions of width $q_{\text{min}}=2\pi/L$, where $L$ is the image size, the number $N_q$ of independent contributions is given by $N_q \simeq \pi q/q_{\text{min}}$. We can thus estimate the uncertainty $\sigma_d$ associated with the measured structure function as
\begin{equation}
\label{sigmad}
    \sigma_d^2 = \frac{1}{N_{\Delta t}N_{q}}(d_2(q,\Delta t)-d(q,\Delta t)^2)\simeq \frac{B^2}{N_{\Delta t}N_{q}}.
\end{equation}
A simple criterion to predict the ability of DDM to reliably measure the dynamics of the sample at a given $q$ (provided that the acquisition frame rate $f_0$ is large enough to correctly sample the relation dynamics, namely $f_0 \gg Dq^2$) corresponds to the requirement that the expected overall variation over time of the image structure function is at least three times larger than the associated errorbar $\sqrt{2}\sigma_d$,  $A>3\sqrt{2}\sigma_d$. Using Eq. \ref{sigmad}, this condition can be written as
\begin{equation}
\label{ineq}
    \text{SNR} > \text{SNR}_{\text{c}},
\end{equation}
where 
\begin{equation}
\label{SNRC}
    \text{SNR}_{\text{c}}=\frac{3\sqrt{2}}{\sqrt{N_{\Delta t}N_{q}}}     
\end{equation}
is the critical signal-to-noise ratio. We can use Eq. \ref{SNRC} to estimate $\text{SNR}_\text{c}$ for the measurements presented in this article using $N_{\Delta t} \simeq 2 \cdot 10^4$,
$N_{q}\simeq 3 \cdot 10^2$ for $q=1$ $\mu$m. We get  $\text{SNR}_{\text{c}} \simeq 1\cdot 10^{-3}$, a value that is rather close to the signal-to-noise ratio corresponding to the lowest measured concentrations for all the proteins considered in this work.
Let us now consider the key factors that determine the values of $A$ and $B$ and therefore of $\text{SNR}$.
In a typical working condition for spatially coherent bright-field imaging, a strong transmitted beam is present, and the optical signal in the images is mainly due to the interference between the transmitted field and the field scattered by  the molecules \cite{giavazzi2009scattering}. In this condition, the mean square amplitude of the signal $A$ is proportional to the square of the average intensity (grayscale value) $I_0$ recorded by the camera sensor. 
On the other hand, since the main contribution to the noise comes from the shot noise, \textit{i.e.}, from the Poissonian fluctuation in the number of collected photoelectrons, the noise mean square amplitude $B$ is expected to scale linearly with  $I_0$, with a prefactor that depends on the full well capacity $\text{FWC}$ of the sensor pixel (in our case corresponding to about 30000 electrons per pixel)
\begin{equation*}
B=I_\text{max}/\text{FWC} \cdot I_0.    
\end{equation*}
In the last equation, $I_\text{max}$ is the grayscale value corresponding to saturation.
In our measurements, the average grayscale value was about one-third of the saturation value $I_0\simeq I_\text{max}/3$, as can be seen from the intensity histograms reported in Figure \ref{fig:Histograms_raw}.\\
Let us now consider a suspension of diluted molecules of molecular weight $M$ at concentration $c$.
If we assume fixed values for the density and for d$n$/d$c$, in the Rayleigh regime, the scattered intensity is proportional to the product $c\cdot M$.
Another important factor that affects the amplitude $A(q)$, and which is not easy to accurately predict \textit{a priori}, is the optical transfer function $T(q)$ of the imaging system.
The optical transfer function depends on the coherence of the illumination beam and the numerical aperture of the collection optics and incorporates also the effect of the finite thickness $h$ of the sample, which can be relevant in particular for low $q$\cite{giavazzi2009scattering}.
Combining all the aforementioned contributions, we obtain the following expression for the signal-to-noise ratio
\begin{equation}
\label{SNR}
\text{SNR}\approx a \cdot (\text{FWC}\cdot I_0/I_\text{max})\cdot c \cdot M,
\end{equation}
where the prefactor $a$ depends on the optical transfer function and the optical contrast (d$n$/d$c$) of the molecules. For our setup, $a$ can be estimated by measuring $\text{SNR}$ for a given combination of experimental parameters.
Let us consider, for example, the most diluted BSA sample ($c=1$ mg mL$^{-1}$, $M=6.6\cdot10^4$ Da). In this case, for $q=0.7$ $\mu$m$^{-1}$, we obtain $\textit{SNR}\simeq 2 \cdot 10^{-3}$ (see also Figure 3(a)). This provides the estimate $a\approx 3\cdot 10^{-12}$ mL mg$^{-1}$Da$^{-1}$.
Although this value is not expected to be of general validity, it could provide an order-of-magnitude reference for experiments performed under similar conditions.
The combination of Eqs. \ref{ineq}, \ref{SNRC}, and \ref{SNR} represents the main result of this section, as it specifies the role of all relevant experimental parameters in determining the feasibility of the measurement and enables the prediction of DDM performance under prescribed conditions.}

\section{\REV{Supplementary Movies}}
\REV{For each protein, we provide two movies showing an example of image acquisition at the highest ($c_1$) and the lowest ($c_4$) concentration investigated, as listed below.
Each movie shows the first 500 frames of the image acquisition, before (top) and after (bottom) background removal. Frames are displayed at a playback speed of 50 fps.
The corresponding intensity histograms are shown in Figures \ref{fig:Histograms_raw} and \ref{fig:Histograms_BGrmvd}.\\
List of supplementary movies:\\
- \textbf{SM1}: BSA sample at concentration $c=c_1$\\
- \textbf{SM2}: BSA sample at concentration $c=c_4$\\
- \textbf{SM3}: Lysozyme sample at concentration $c=c_1$\\
- \textbf{SM4}: Lysozyme sample at concentration $c=c_4$\\
- \textbf{SM5}: Pepsin sample at concentration $c=c_1$\\
- \textbf{SM6}: Pepsin sample at concentration $c=c_4$\\}

\begin{figure*}[h]
\centering
\includegraphics[width=\textwidth]{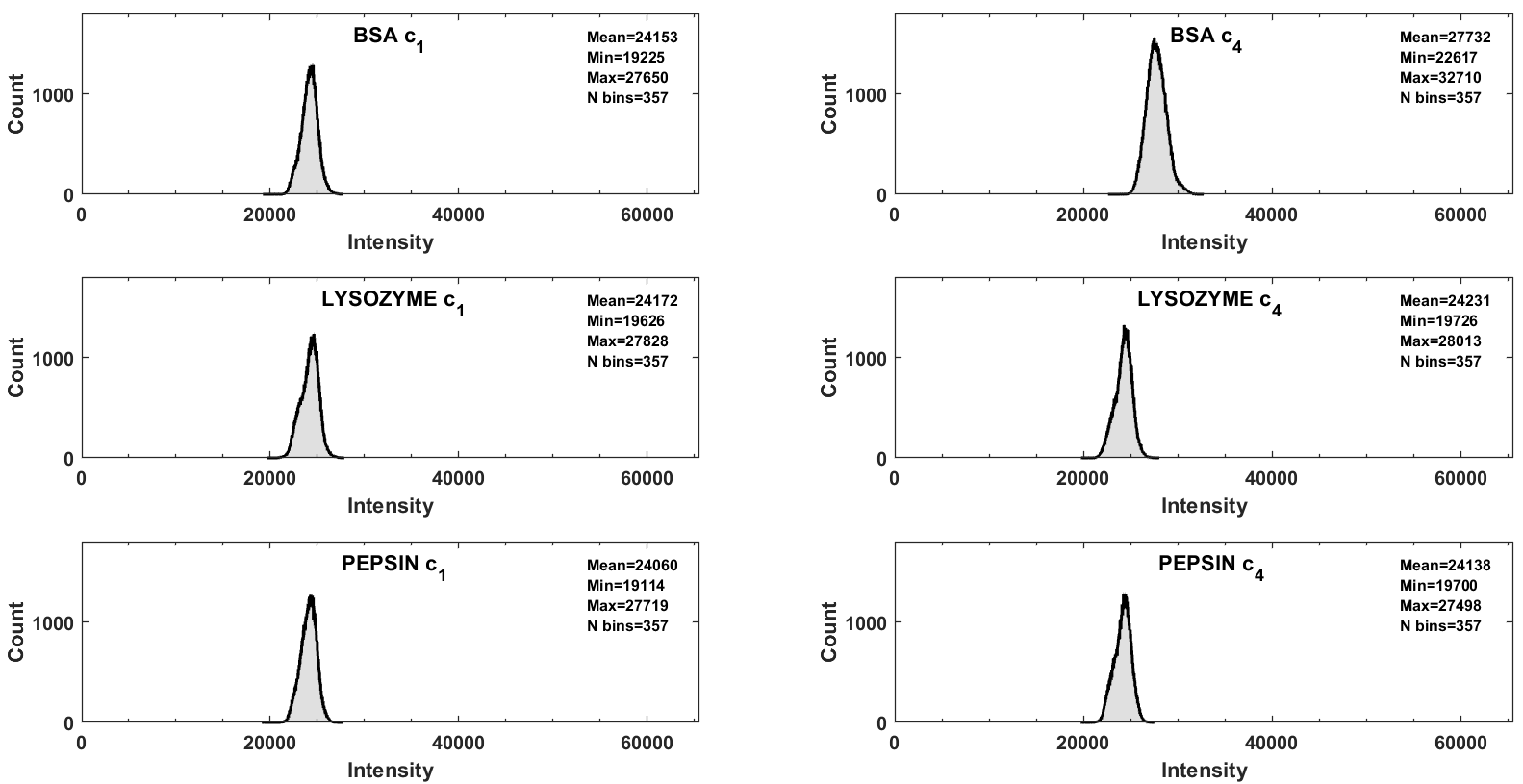}
\caption{\REV{\textbf{Examples of intensity histograms.} Intensity histograms of the raw images shown in the Supplementary Movies SM1-SM6.}}
\label{fig:Histograms_raw}
\end{figure*}

\begin{figure*}[h]
\centering
\includegraphics[width=\textwidth]{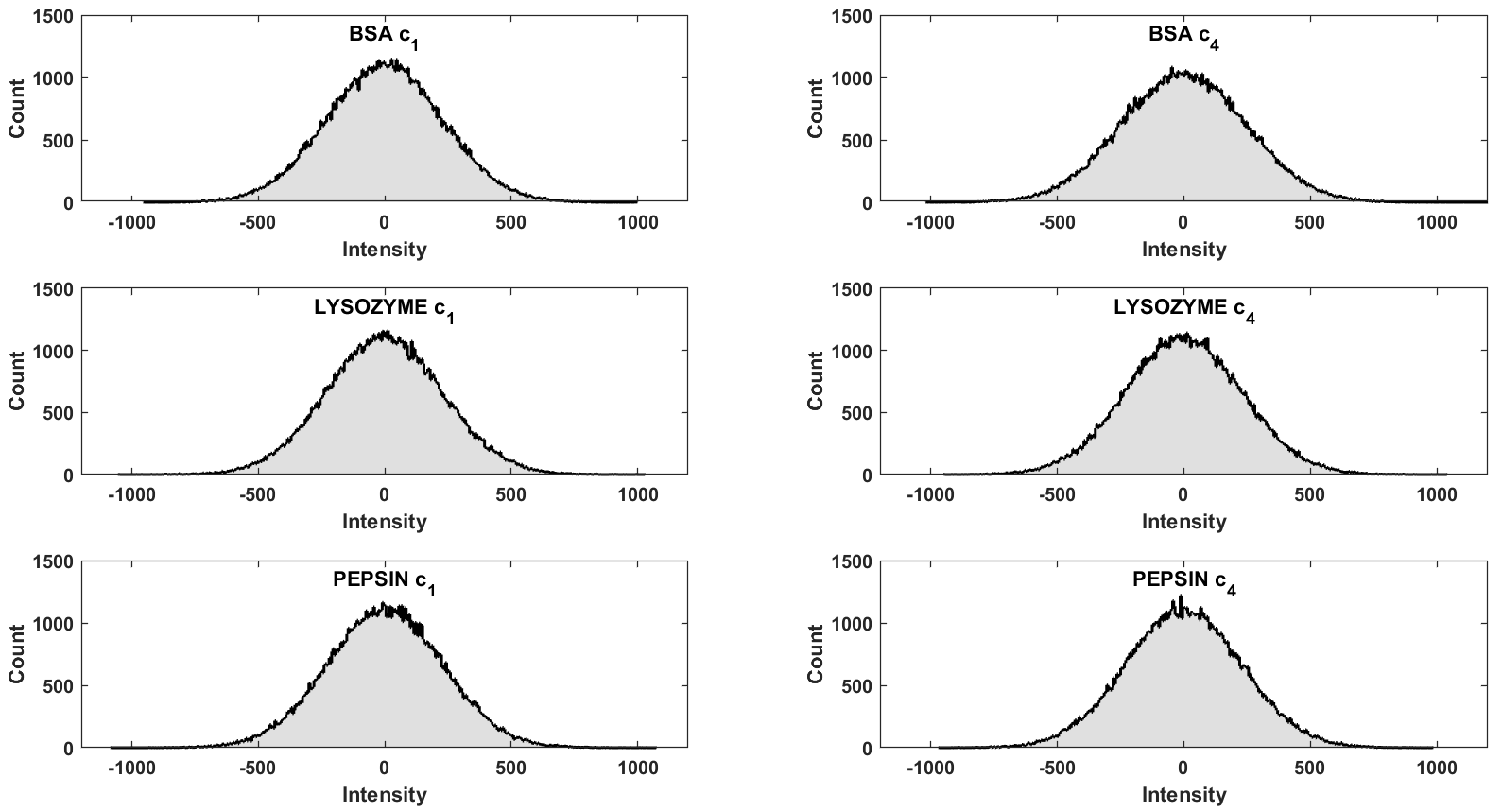}
\caption{\REV{\textbf{Examples of intensity histograms after background subtraction.} Intensity histograms of the background-subtracted images shown in the Supplementary Movies SM1-SM6.}}
\label{fig:Histograms_BGrmvd}
\end{figure*}

\section{\REV{Supplementary Data}}
\begin{figure*}[h]
\centering
\includegraphics[width=\textwidth]{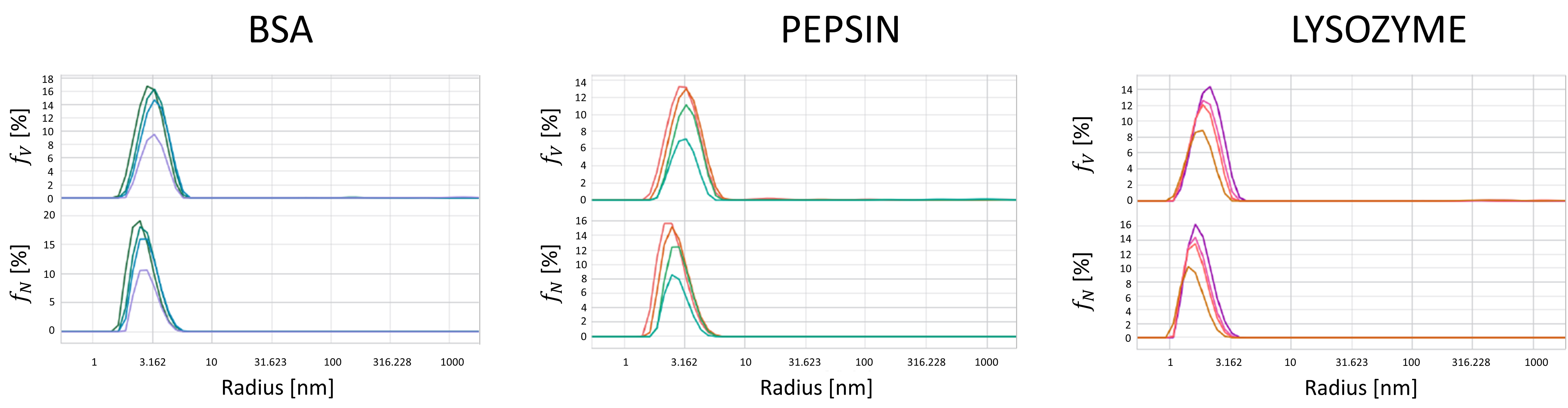}
\caption{\textbf{Protein size distributions.} Volume-weighted $f_V$ (top row) and number-weighted $f_N$ (bottom row) size distributions obtained with DLS for the different proteins.}
\label{fig:DLS_dist}
\end{figure*}

\begin{figure*}[h]
\centering
\includegraphics[width=\textwidth]{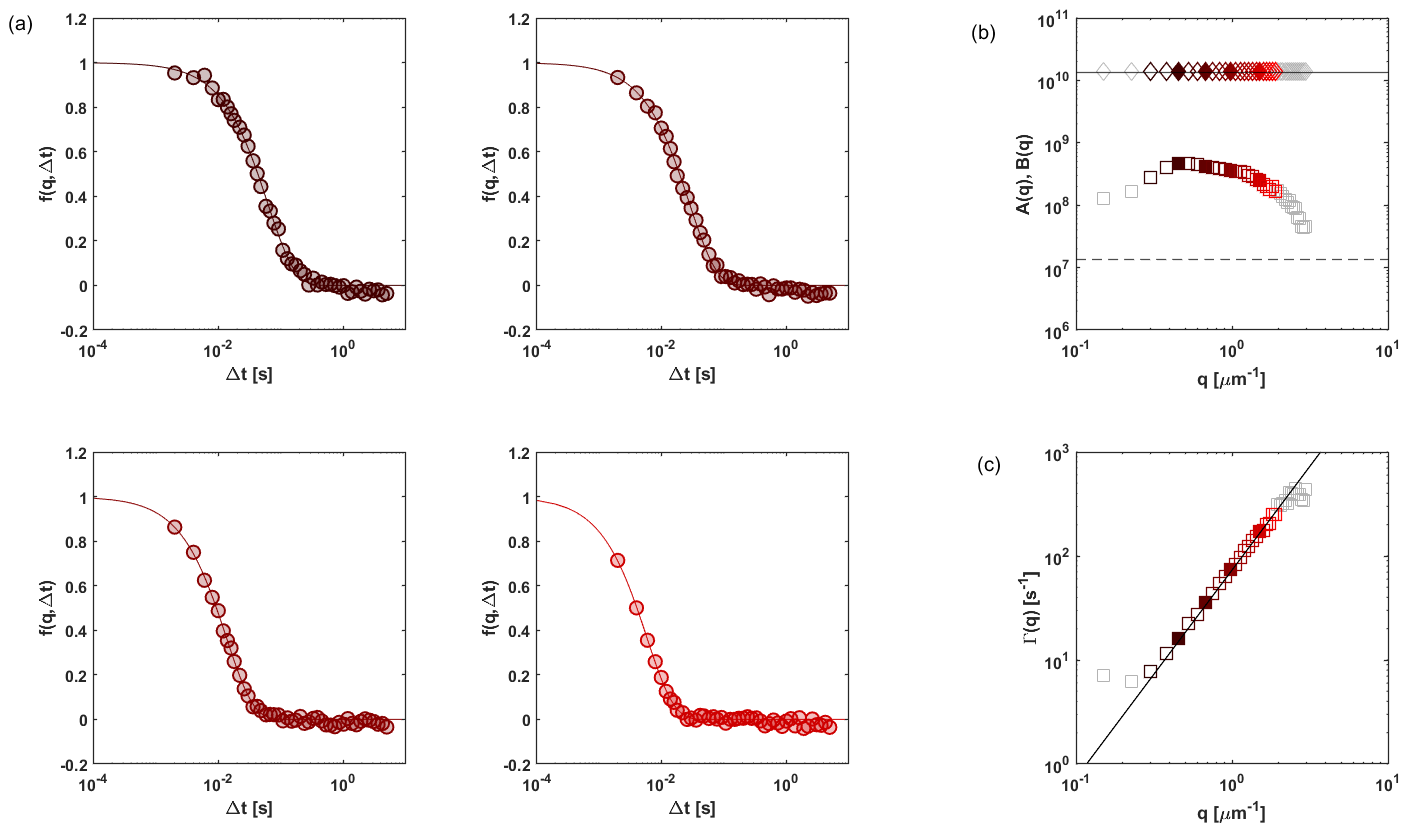}
\caption{\textbf{BSA sample at $c=c_1$.} (a) Examples of intermediate scattering functions $f(q,\Delta t)$ at different wavevectors $q$. From top left to bottom right: $q$=0.45 $\mu$m$^{-1}$, 0.68 $\mu$m$^{-1}$, 0.98 $\mu$m$^{-1}$, 1.51 $\mu$m$^{-1}$. Circles correspond to experimental data, solid lines correspond to a single exponential fit of the kind $f(q,\Delta t)=\exp(-\Gamma(q) \Delta t)$. (b) Signal amplitude $A(q)$ (squares) and noise term $B(q)$ (diamonds). The black solid line represents the average noise level, while the dashed line marks a SNR equal to $10^{-3}$. (c) Relaxation rates $\Gamma(q)$. Colored squares mark the range of $q$ considered for the quadratic fit.
%, in which the SNR is at least 1/300 and the relaxation rate does not exceed half the frame rate, namely $\Gamma(q)<250$ s$^{-1}$.
The weighted quadratic fit, shown as a black solid line, gives a diffusion coefficient $D$ equal to {(73 $\pm$ 2) $\mu$m$^2$/s}. Solid symbols in (b) and (c) highlight the wavevectors $q$ for which the decay of $f(q,\Delta t)$ is shown in panel (a).}
\label{fig:BSA_A1}
\end{figure*}

\begin{figure*}[h]
\centering
\includegraphics[width=\textwidth]{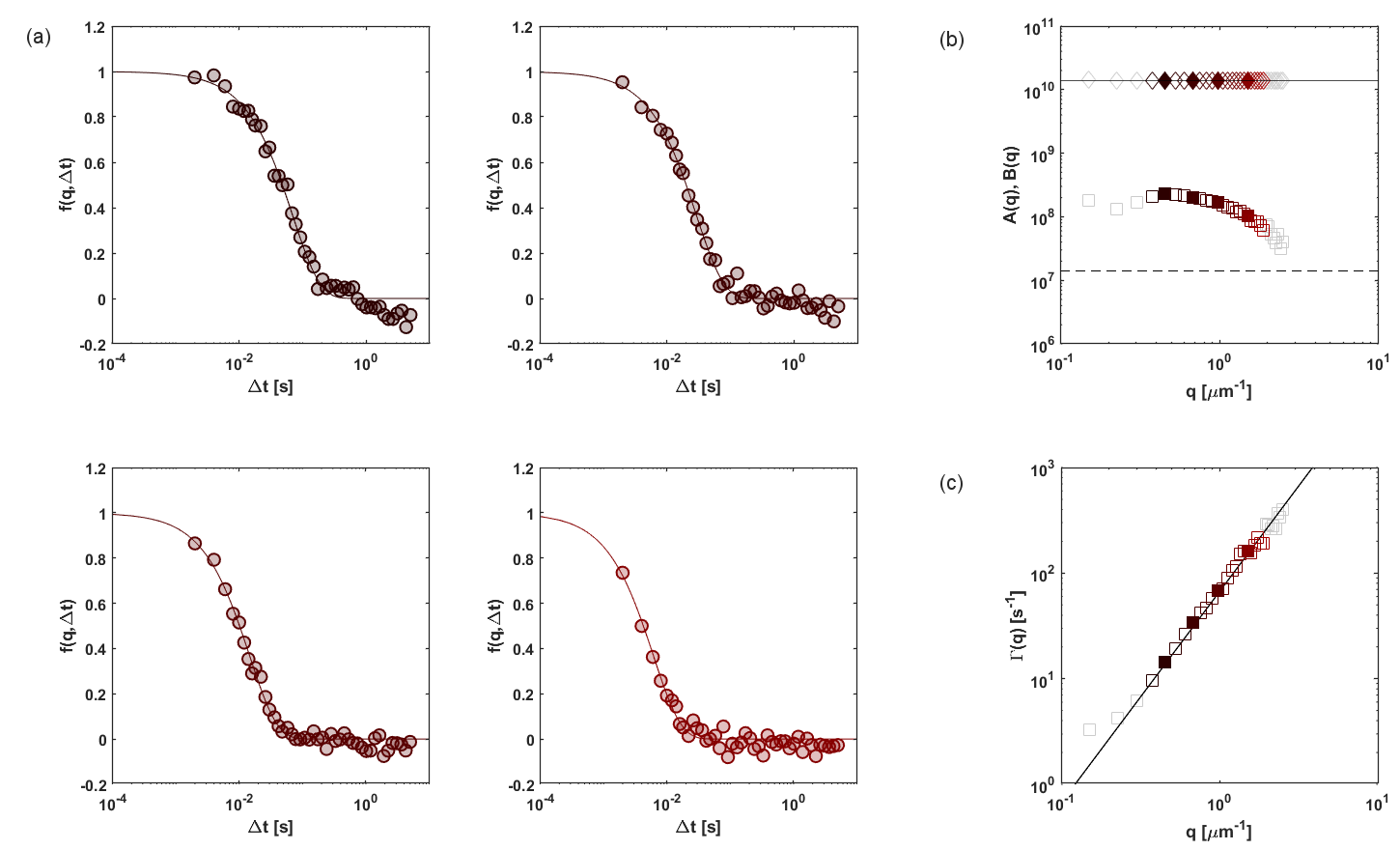}
\caption{\textbf{BSA sample at $c=c_2$.} (a) Examples of intermediate scattering functions $f(q,\Delta t)$ at different wavevectors $q$. From top left to bottom right: $q$=0.45 $\mu$m$^{-1}$, 0.68 $\mu$m$^{-1}$, 0.98 $\mu$m$^{-1}$, 1.51 $\mu$m$^{-1}$. Circles correspond to experimental data, solid lines correspond to a single exponential fit of the kind $f(q,\Delta t)=\exp(-\Gamma(q) \Delta t)$. (b) Signal amplitude $A(q)$ (squares) and noise term $B(q)$ (diamonds). The black solid line represents the average noise level, while the dashed line marks a SNR equal to $10^{-3}$. (c) Relaxation rates $\Gamma(q)$. Colored squares mark the range of $q$ considered for the quadratic fit. The weighted quadratic fit, shown as a black solid line, gives a diffusion coefficient $D$ equal to (67 $\pm$ 4) $\mu$m$^2$/s. Solid symbols in (b) and (c) highlight the wavevectors $q$ for which the decay of $f(q,\Delta t)$ is shown in panel (a).}
\label{fig:BSA_A2}
\end{figure*}

\begin{figure*}[h]
\centering
\includegraphics[width=\textwidth]{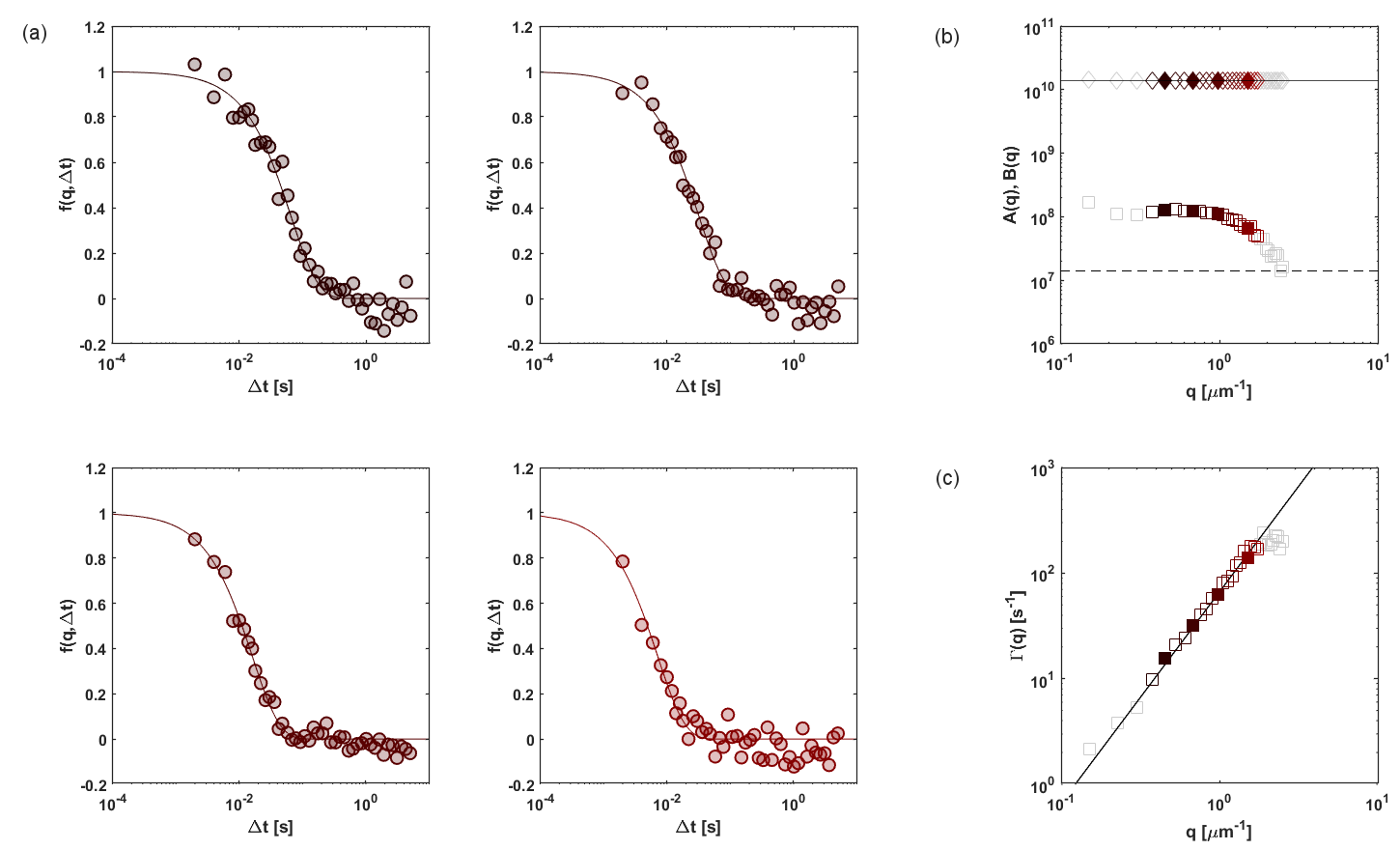}
\caption{\textbf{BSA sample at $c=c_3$.} (a) Examples of intermediate scattering functions $f(q,\Delta t)$ at different wavevectors $q$. From top left to bottom right: $q$=0.45 $\mu$m$^{-1}$, 0.68 $\mu$m$^{-1}$, 0.98 $\mu$m$^{-1}$, 1.51 $\mu$m$^{-1}$. Circles correspond to experimental data, solid lines correspond to a single exponential fit of the kind $f(q,\Delta t)=\exp(-\Gamma(q) \Delta t)$. (b) Signal amplitude $A(q)$ (squares) and noise term $B(q)$ (diamonds). The black solid line represents the average noise level, while the dashed line marks a SNR equal to $10^{-3}$. (c) Relaxation rates $\Gamma(q)$. Colored squares mark the range of $q$ considered for the quadratic fit. The weighted quadratic fit, shown as a black solid line, gives a diffusion coefficient $D$ equal to (66 $\pm$ 3) $\mu$m$^2$/s. Solid symbols in (b) and (c) highlight the wavevectors $q$ for which the decay of $f(q,\Delta t)$ is shown in panel (a).}
\label{fig:BSA_A3.png}
\end{figure*}

\begin{figure*}[h]
\centering
\includegraphics[width=\textwidth]{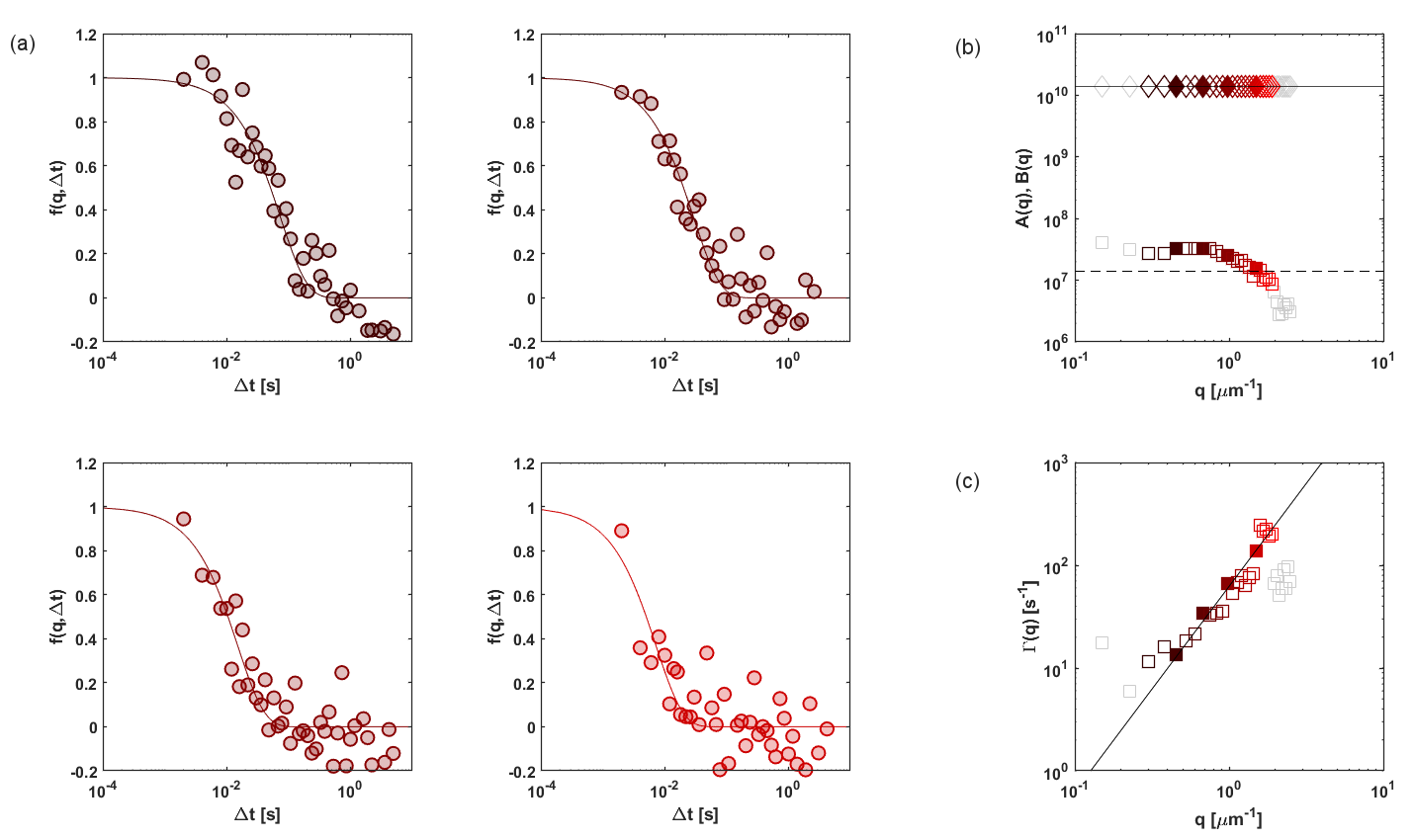}
\caption{\textbf{BSA sample at $c=c_4$.} (a) Examples of intermediate scattering functions $f(q,\Delta t)$ at different wavevectors $q$. From top left to bottom right: $q$=0.45 $\mu$m$^{-1}$, 0.68 $\mu$m$^{-1}$, 0.98 $\mu$m$^{-1}$, 1.51 $\mu$m$^{-1}$. Circles correspond to experimental data, solid lines correspond to a single exponential fit of the kind $f(q,\Delta t)=\exp(-\Gamma(q) \Delta t)$. (b) Signal amplitude $A(q)$ (squares) and noise term $B(q)$ (diamonds). The black solid line represents the average noise level, while the dashed line marks a SNR equal to $10^{-3}$. (c) Relaxation rates $\Gamma(q)$. Colored squares mark the range of $q$ considered for the quadratic fit. The weighted quadratic fit, shown as a black solid line, gives a diffusion coefficient $D$ equal to (62 $\pm$ 7) $\mu$m$^2$/s. Solid symbols in (b) and (c) highlight the wavevectors $q$ for which the decay of $f(q,\Delta t)$ is shown in panel (a).}
\label{fig:BSA_A4_New.png}
\end{figure*}

\begin{figure*}[h]
\centering
\includegraphics[width=\textwidth]{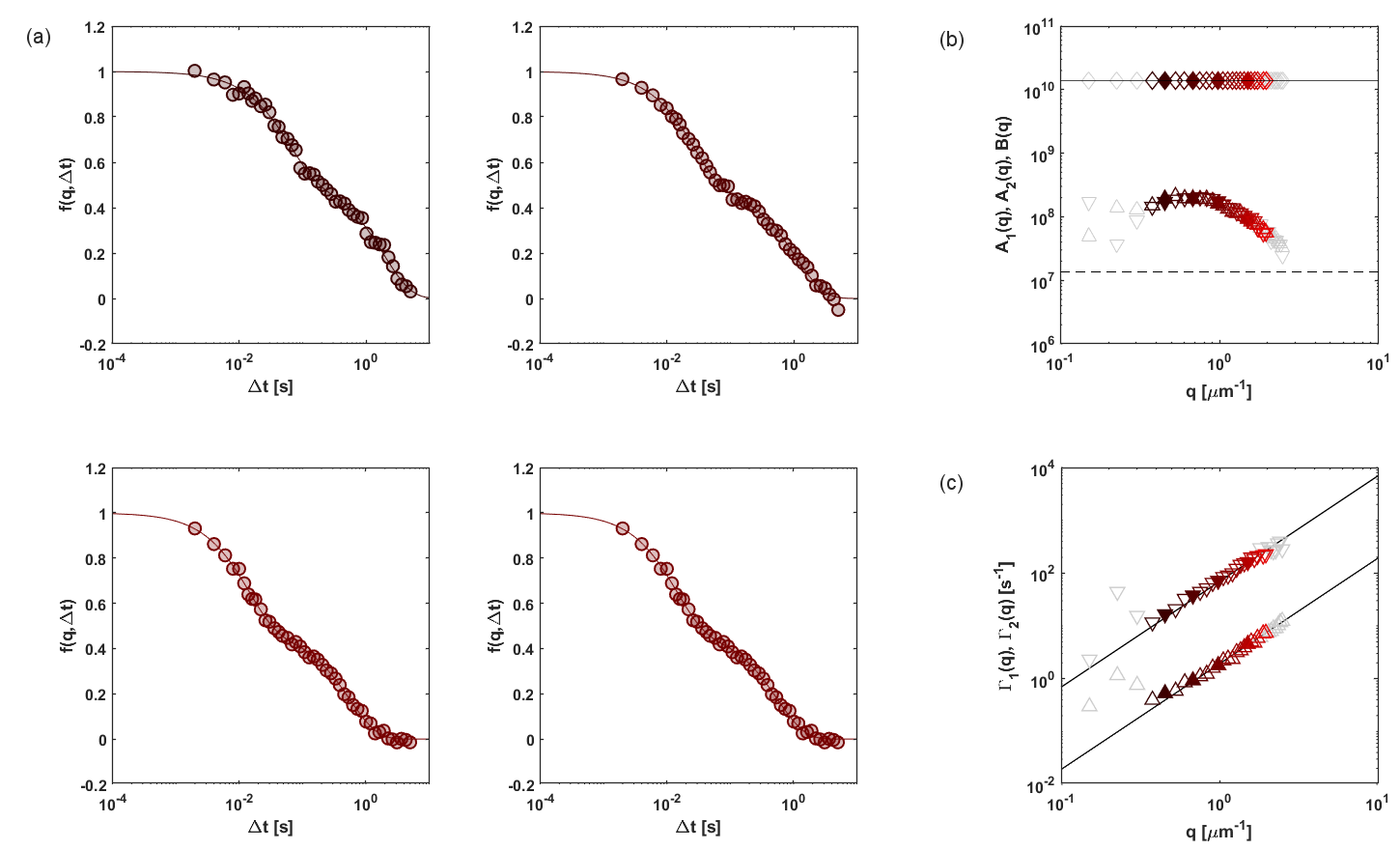}
\caption{\textbf{Pepsin sample at  $c=c_1$.} (a) Intermediate scattering functions $f(q,\Delta t)$ at different wavevectors $q$. From top left to bottom right: $q$=0.45 $\mu$m$^{-1}$, 0.68 $\mu$m$^{-1}$, 0.98 $\mu$m$^{-1}$, 1.51 $\mu$m$^{-1}$. Circles correspond to experimental data, solid lines correspond to a double exponential fit of the kind $f(q,\Delta t)=\alpha_1 \exp(-\Gamma_1(q) \Delta t) + \alpha_2 \exp(-\Gamma_2(q) \Delta t)$. (b) Signal amplitudes $A_1(q) = \alpha_1\cdot A$ (down-pointed triangles) and $A_2(q) = \alpha_2 \cdot A$ (up-pointed triangles), and noise term $B(q)$ (diamonds).  The black solid line represents the average noise level, while the dashed line marks a SNR equal to $10^{-3}$. (c) Relaxation rates $\Gamma_1(q)$ (down-pointed triangles) and $\Gamma_2(q)$ (up-pointed triangles). Colored triangles mark the range of $q$ considered for the quadratic fits.
The two quadratic fits, shown as black solid lines, give diffusion coefficients $D_1$=(69 $\pm$ 3) $\mu$m$^2$/s and $D_2$=(1.92 $\pm$ 0.05) $\mu$m$^2$/s. Solid symbols in (b) and (c) highlight the wavevectors $q$ for which the decay of $f(q,\Delta t)$ is shown in panel (a).}
\label{fig:PEP_B1}
\end{figure*}

\begin{figure*}[h]
\centering
\includegraphics[width=\textwidth]{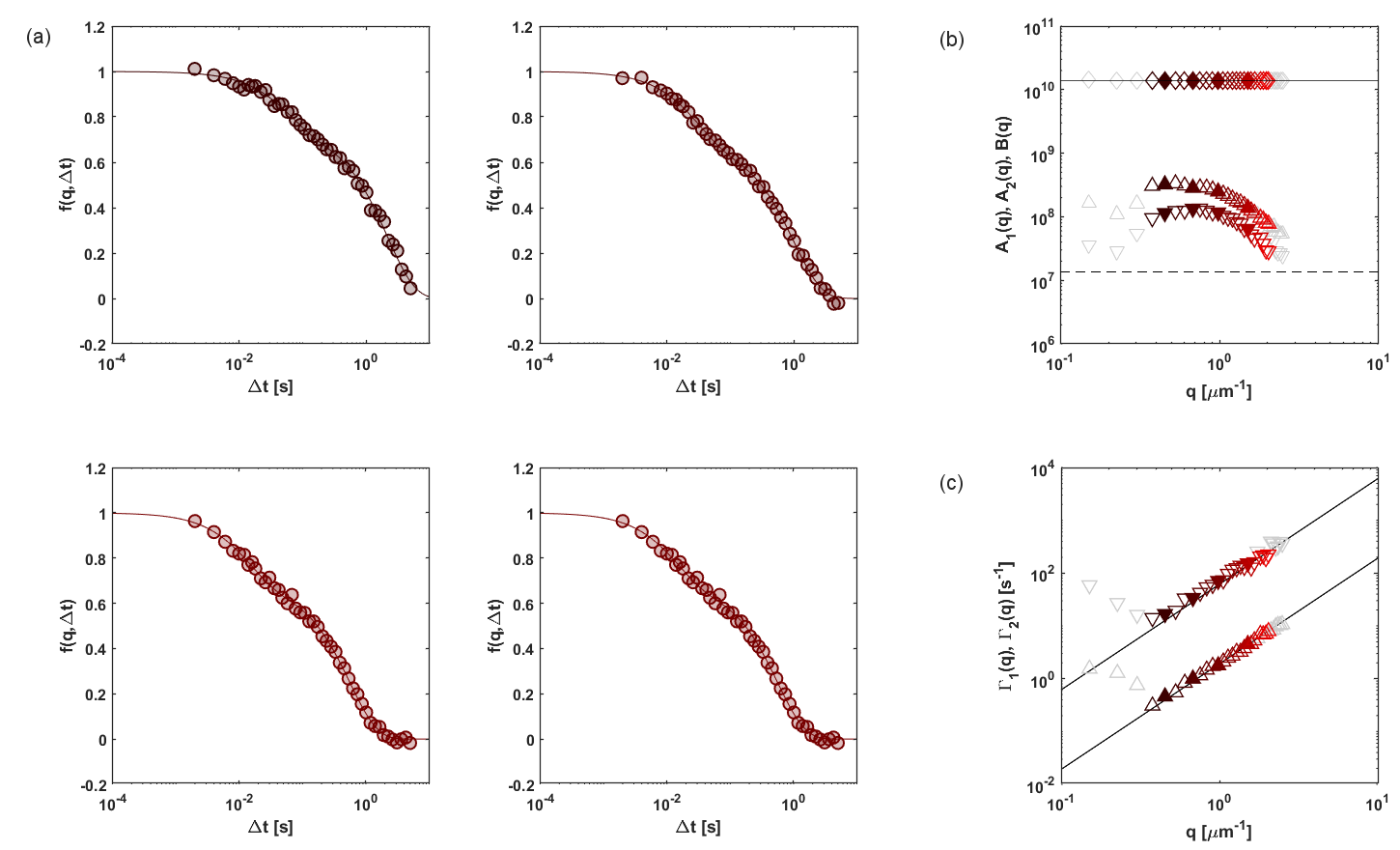}
\caption{\textbf{Pepsin sample at $c=c_2$.} (a) Intermediate scattering functions $f(q,\Delta t)$ at different wavevectors $q$. From top left to bottom right: $q$=0.45 $\mu$m$^{-1}$, 0.68 $\mu$m$^{-1}$, 0.98 $\mu$m$^{-1}$, 1.51 $\mu$m$^{-1}$. Circles correspond to experimental data, solid lines correspond to a double exponential fit of the kind $f(q,\Delta t)=\alpha_1 \exp(-\Gamma_1(q) \Delta t) + \alpha_2 \exp(-\Gamma_2(q) \Delta t)$. (b) Signal amplitudes $A_1(q) = \alpha_1\cdot A$ (down-pointed triangles) and $A_2(q) = \alpha_2 \cdot A$ (up-pointed triangles), and noise term $B(q)$ (diamonds).  The black solid line represents the average noise level, while the dashed line marks a SNR equal to $10^{-3}$. (c) Relaxation rates $\Gamma_1(q)$ (down-pointed triangles) and $\Gamma_2(q)$ (up-pointed triangles). Colored triangles mark the range of $q$ considered for the quadratic fits. The two quadratic fits, shown as black solid lines, give diffusion coefficients $D_1$=(61 $\pm$ 4) $\mu$m$^2$/s and $D_2$=(1.92 $\pm$ 0.05) $\mu$m$^2$/s. Solid symbols in (b) and (c) highlight the wavevectors $q$ for which the decay of $f(q,\Delta t)$ is shown in panel (a).}
\label{fig:PEP_B2}
\end{figure*}

\begin{figure*}[h]
\centering
\includegraphics[width=\textwidth]{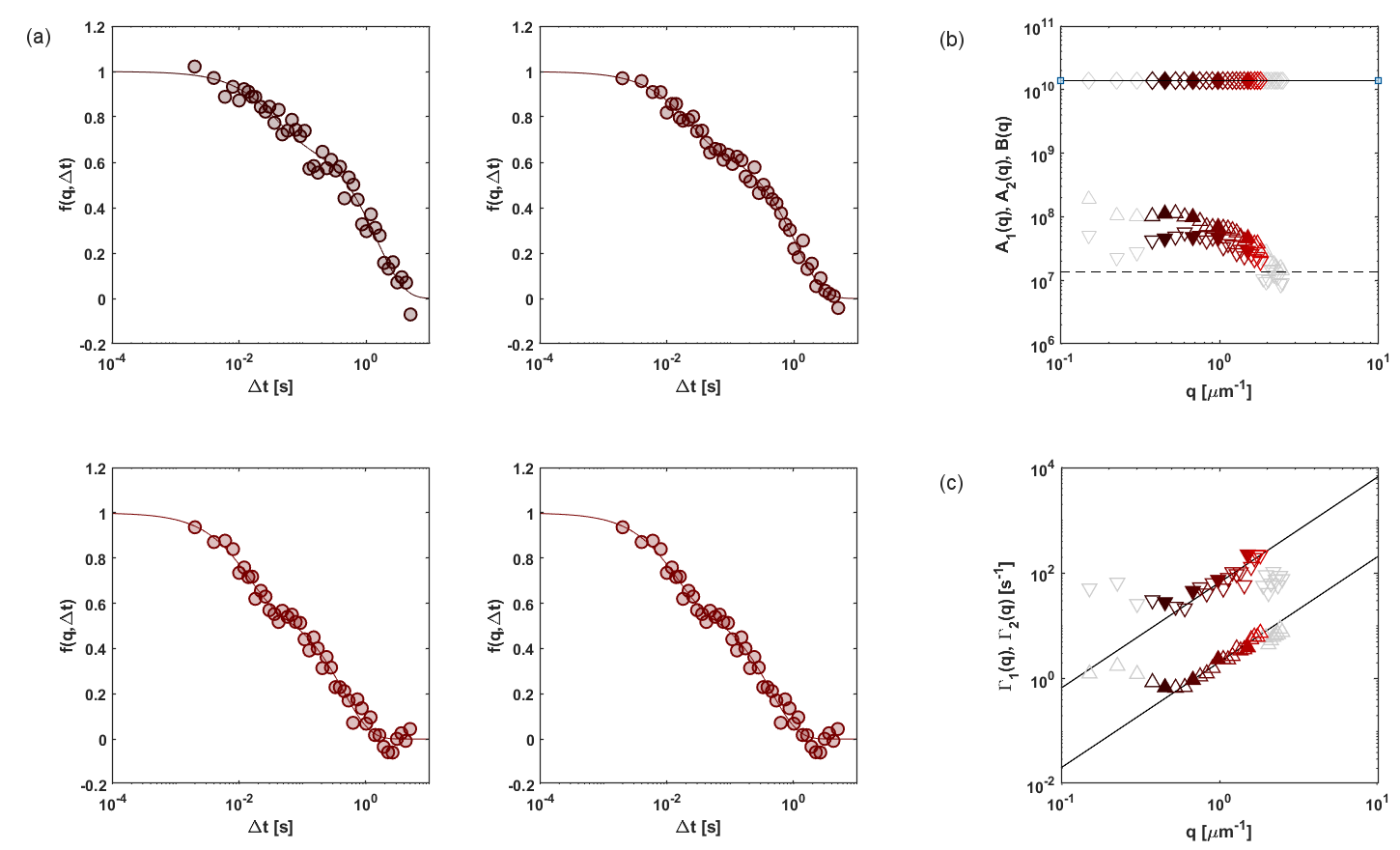}
\caption{\textbf{Pepsin sample at $c=c_3$.} (a) Intermediate scattering functions $f(q,\Delta t)$ at different wavevectors $q$. From top left to bottom right: $q$=0.45 $\mu$m$^{-1}$, 0.68 $\mu$m$^{-1}$, 0.98 $\mu$m$^{-1}$, 1.51 $\mu$m$^{-1}$. Circles correspond to experimental data, solid lines correspond to a double exponential fit of the kind $f(q,\Delta t)=\alpha_1 \exp(-\Gamma_1(q) \Delta t) + \alpha_2 \exp(-\Gamma_2(q) \Delta t)$. (b) Signal amplitudes $A_1(q) = \alpha_1\cdot A$ (down-pointed triangles) and $A_2(q) = \alpha_2 \cdot A$ (up-pointed triangles), and noise term $B(q)$ (diamonds).  The black solid line represents the average noise level, while the dashed line marks a SNR equal to $10^{-3}$. (c) Relaxation rates $\Gamma_1(q)$ (down-pointed triangles) and $\Gamma_2(q)$ (up-pointed triangles). Colored triangles mark the range of $q$ considered for the quadratic fits. The two quadratic fits, shown as black solid lines, give diffusion coefficients $D_1$=(66 $\pm$ 9) $\mu$m$^2$/s and $D_2$=(2.1 $\pm$ 0.1) $\mu$m$^2$/s. Solid symbols in (b) and (c) highlight the wavevectors $q$ for which the decay of $f(q,\Delta t)$ is shown in panel (a).}
\label{fig:PEP_B3}
\end{figure*}

\begin{figure*}[h]
\centering
\includegraphics[width=\textwidth]{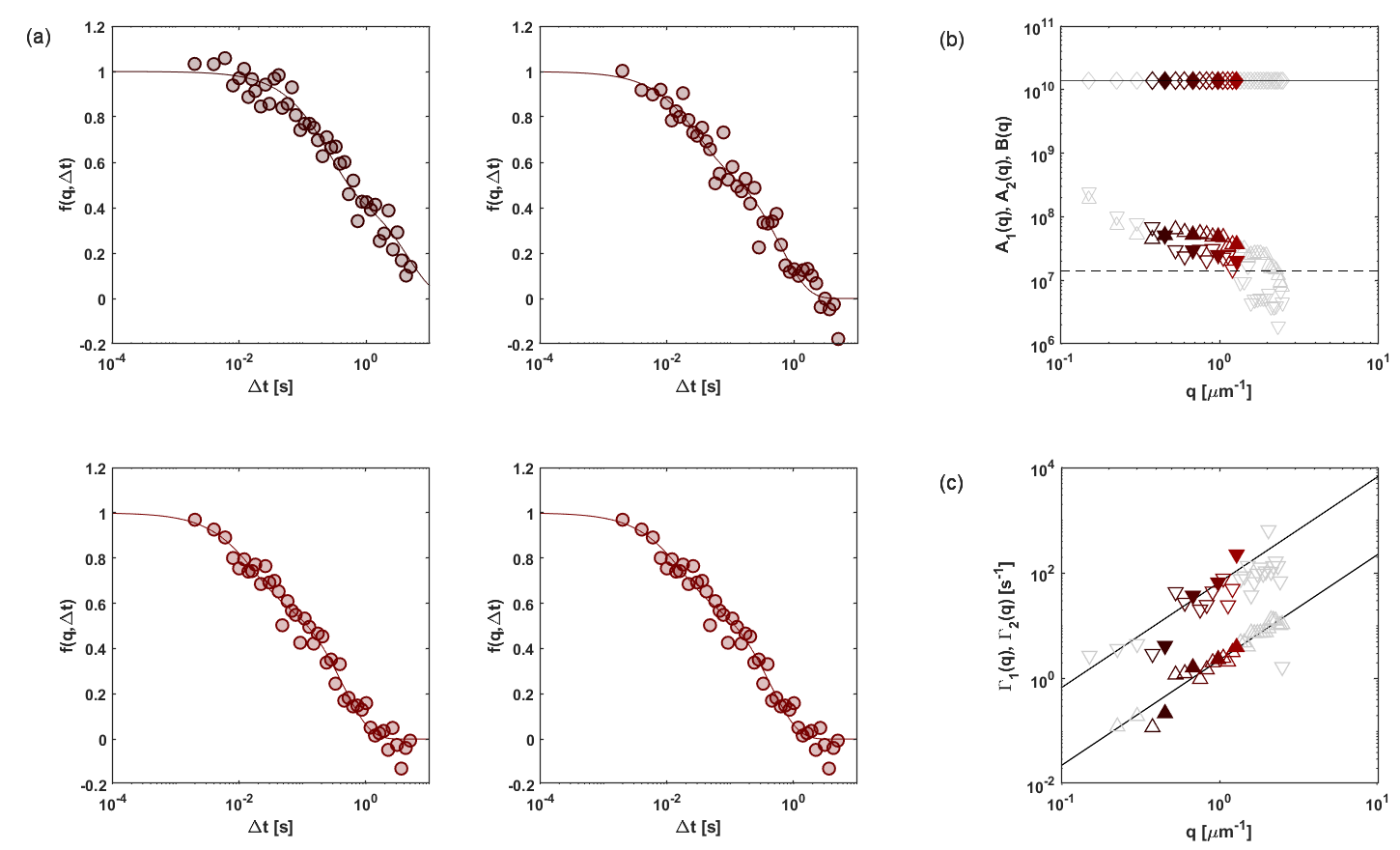}
\caption{\textbf{Pepsin sample at $c=c_4$.} (a) Intermediate scattering functions $f(q,\Delta t)$ at different wavevectors $q$. From top left to bottom right: $q$=0.45 $\mu$m$^{-1}$, 0.68 $\mu$m$^{-1}$, 0.98 $\mu$m$^{-1}$, 1.28 $\mu$m$^{-1}$. Circles correspond to experimental data, solid lines correspond to a double exponential fit of the kind $f(q,\Delta t)=\alpha_1 \exp(-\Gamma_1(q) \Delta t) + \alpha_2 \exp(-\Gamma_2(q) \Delta t)$. (b) Signal amplitudes $A_1(q) = \alpha_1\cdot A$ (down-pointed triangles) and $A_2(q) = \alpha_2 \cdot A$ (up-pointed triangles), and noise term $B(q)$ (diamonds).  The black solid line represents the average noise level $\bar{B}$, while the dashed line marks a SNR equal to $10^{-3}$. (c) Relaxation rates $\Gamma_1(q)$ (down-pointed triangles) and $\Gamma_2(q)$ (up-pointed triangles). Colored triangles mark the range of $q$ considered for the quadratic fits. The two quadratic fits, shown as black solid lines, give diffusion coefficients $D_1$=(67 $\pm$ 27) $\mu$m$^2$/s and $D_2$=(2.3 $\pm$ 0.3) $\mu$m$^2$/s. Solid symbols in (b) and (c) highlight the wavevectors $q$ for which the decay of $f(q,\Delta t)$ is shown in panel (a).}
\label{fig:PEP_B4}
\end{figure*}

\begin{figure*}[h]
\centering
\includegraphics[width=\textwidth]{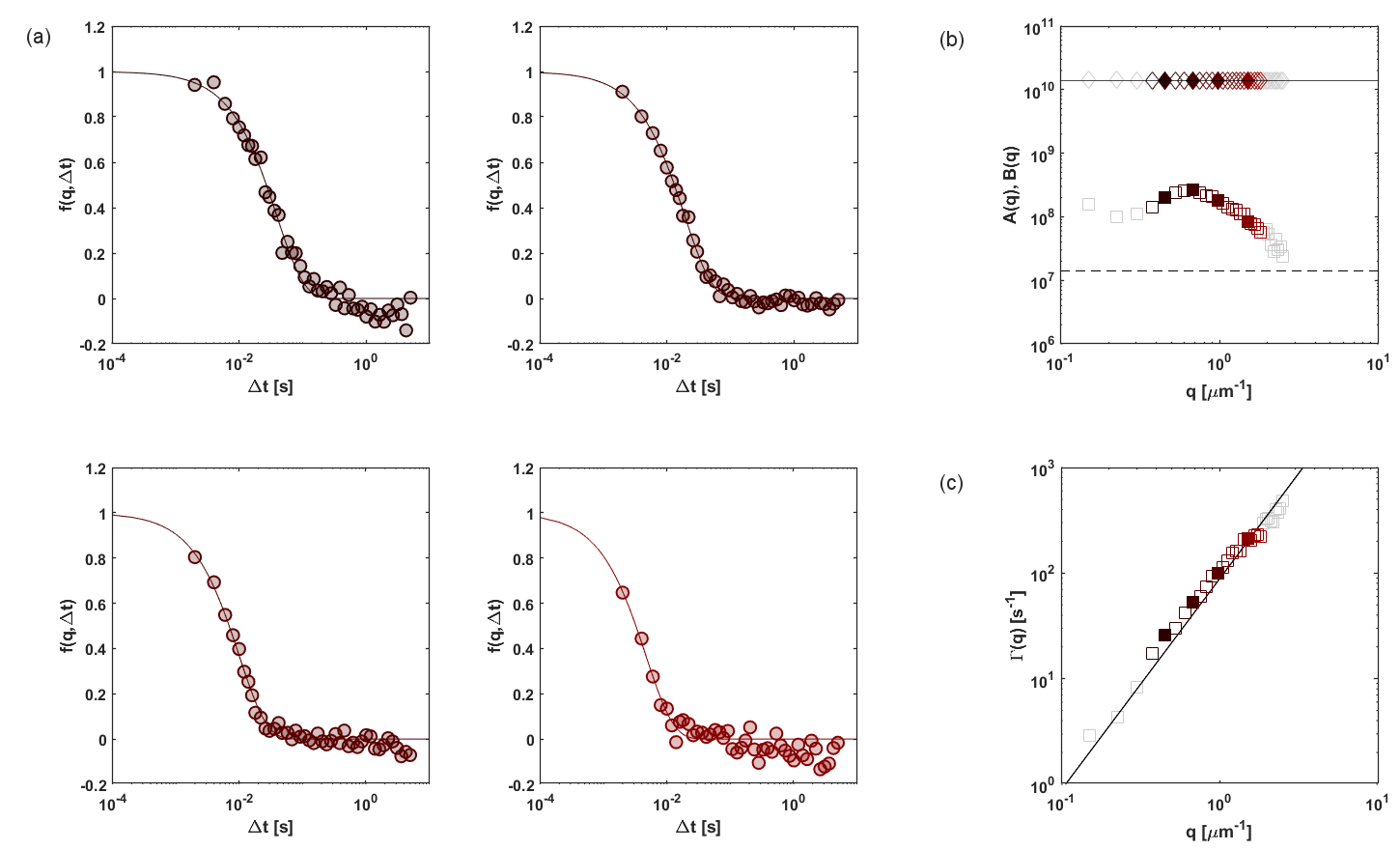}
\caption{\textbf{Lysozyme sample at $c=c_1$.} (a) Intermediate scattering functions $f(q,\Delta t)$ at different wavevectors $q$. From top left to bottom right: $q$=0.45 $\mu$m$^{-1}$, 0.68 $\mu$m$^{-1}$, 0.98 $\mu$m$^{-1}$, 1.51 $\mu$m$^{-1}$. Circles correspond to experimental data, solid lines correspond to a single exponential fit of the kind $f(q,\Delta t)=\exp(-\Gamma(q) \Delta t)$. (b) Signal amplitude $A(q)$ (squares) and noise term $B(q)$ (diamonds). The black solid line represents the average noise level, while the dashed line marks a SNR equal to $10^{-3}$. (c) Relaxation rates $\Gamma(q)$. Colored squares mark the range of $q$ considered for the quadratic fit. The weighted quadratic fit, shown as a black solid line, gives a diffusion coefficient $D$ equal to (88 $\pm$ 6) $\mu$m$^2$/s. Solid symbols in (b) and (c) highlight the wavevectors $q$ for which the decay of $f(q,\Delta t)$ is shown in panel (a).}
\label{fig:LYSO_C1}
\end{figure*}

\begin{figure*}[h]
\centering
\includegraphics[width=\textwidth]{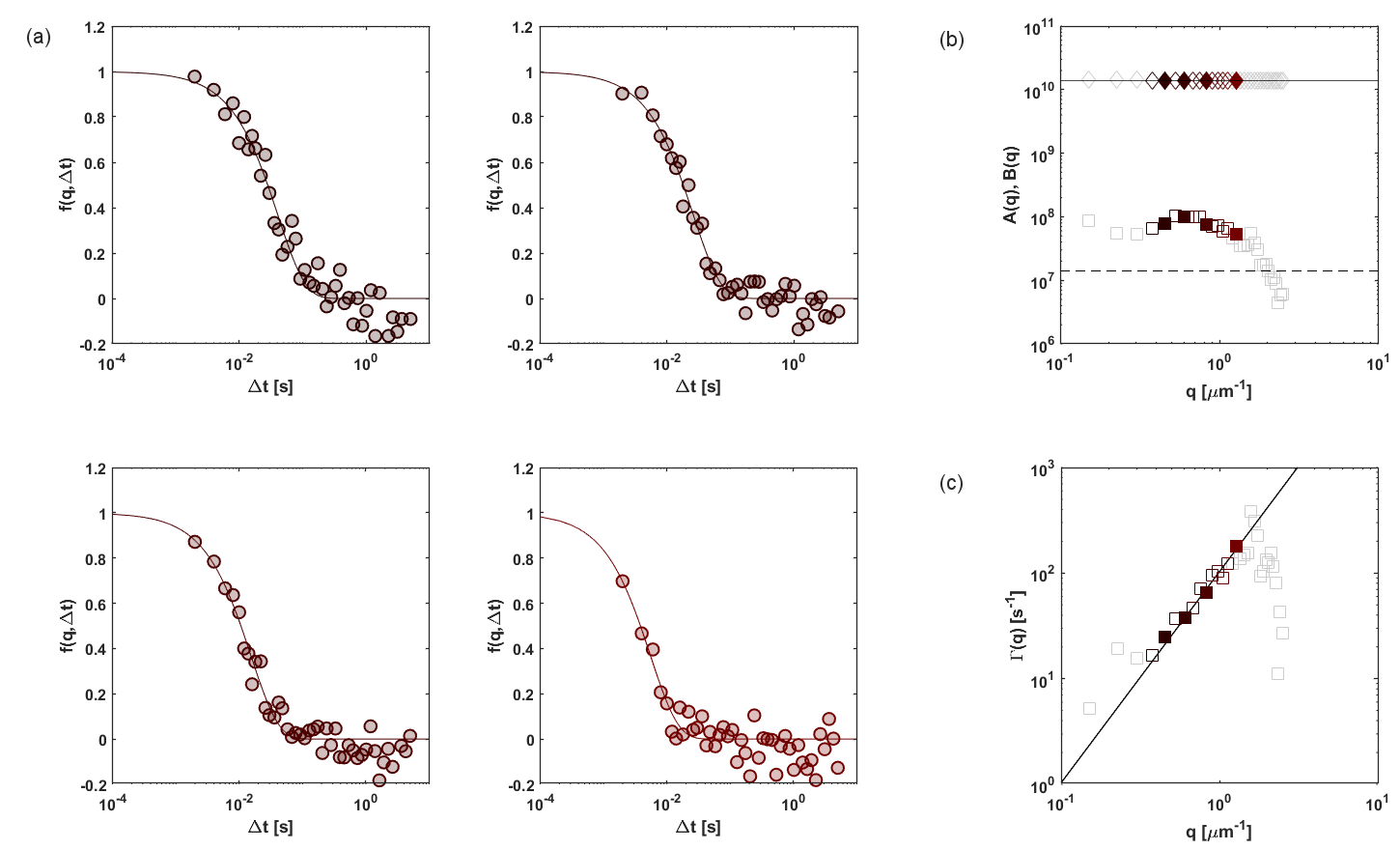}
\caption{\textbf{Lysozyme sample at $c=c_2$.} (a) Intermediate scattering functions $f(q,\Delta t)$ at different wavevectors $q$. From top left to bottom right: $q$=0.45 $\mu$m$^{-1}$, 0.60 $\mu$m$^{-1}$, 0.83 $\mu$m$^{-1}$, 1.28 $\mu$m$^{-1}$. Circles correspond to experimental data, solid lines correspond to a single exponential fit of the kind $f(q,\Delta t)=\exp(-\Gamma(q) \Delta t)$. (b) Signal amplitude $A(q)$ (squares) and noise term $B(q)$ (diamonds). The black solid line represents the average noise level, while the dashed line marks a SNR equal to $10^{-3}$. (c) Relaxation rates $\Gamma(q)$. Colored squares mark the range of $q$ considered for the quadratic fit. The weighted quadratic fit, shown as a black solid line, gives a diffusion coefficient $D$ equal to (103 $\pm$ 8) $\mu$m$^2$/s. Solid symbols in (b) and (c) highlight the wavevectors $q$ for which the decay of $f(q,\Delta t)$ is shown in panel (a).}
\label{fig:LYSO_C2}
\end{figure*}

\begin{figure*}[h]
\centering
\includegraphics[width=\textwidth]{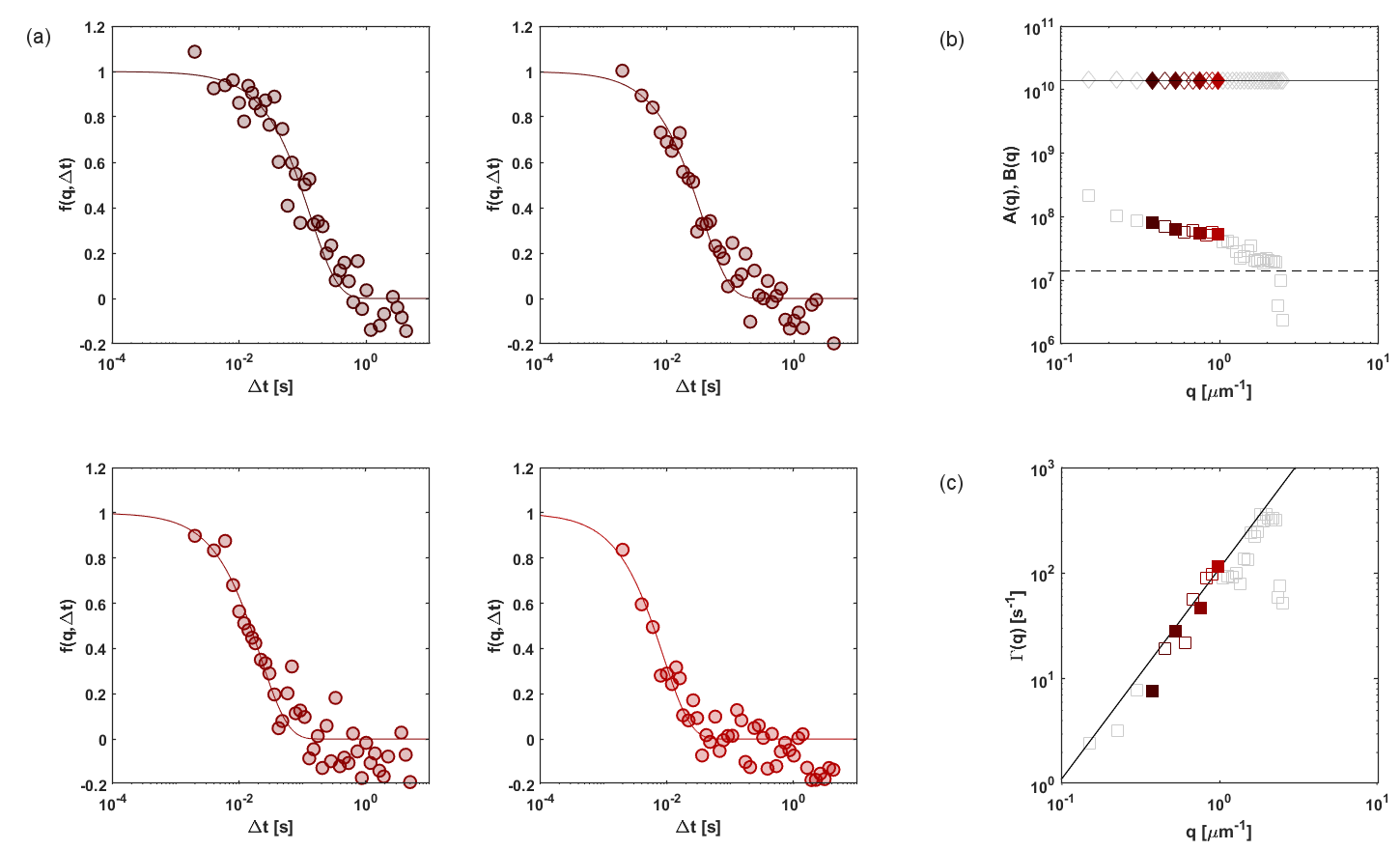}
\caption{\textbf{Lysozyme sample at $c=c_3$.} (a) Intermediate scattering functions $f(q,\Delta t)$ at different wavevectors $q$. From top left to bottom right: $q$=0.38 $\mu$m$^{-1}$, 0.53 $\mu$m$^{-1}$, 0.76 $\mu$m$^{-1}$, 0.98 $\mu$m$^{-1}$. Circles correspond to experimental data, solid lines correspond to a single exponential fit of the kind $f(q,\Delta t)=\exp(-\Gamma(q) \Delta t)$. (b) Signal amplitude $A(q)$ (squares) and noise term $B(q)$ (diamonds). The black solid line represents the average noise level, while the dashed line marks a SNR equal to $10^{-3}$. (c) Relaxation rates $\Gamma(q)$. Colored squares mark the range of $q$ considered for the quadratic fit. The weighted quadratic fit, shown as a black solid line, gives a diffusion coefficient $D$ equal to (111 $\pm$ 17) $\mu$m$^2$/s. Solid symbols in (b) and (c) highlight the wavevectors $q$ for which the decay of $f(q,\Delta t)$ is shown in panel (a).}
\label{fig:LYSO_C3}
\end{figure*}

\begin{figure*}[h]
\centering
\includegraphics[width=\textwidth]{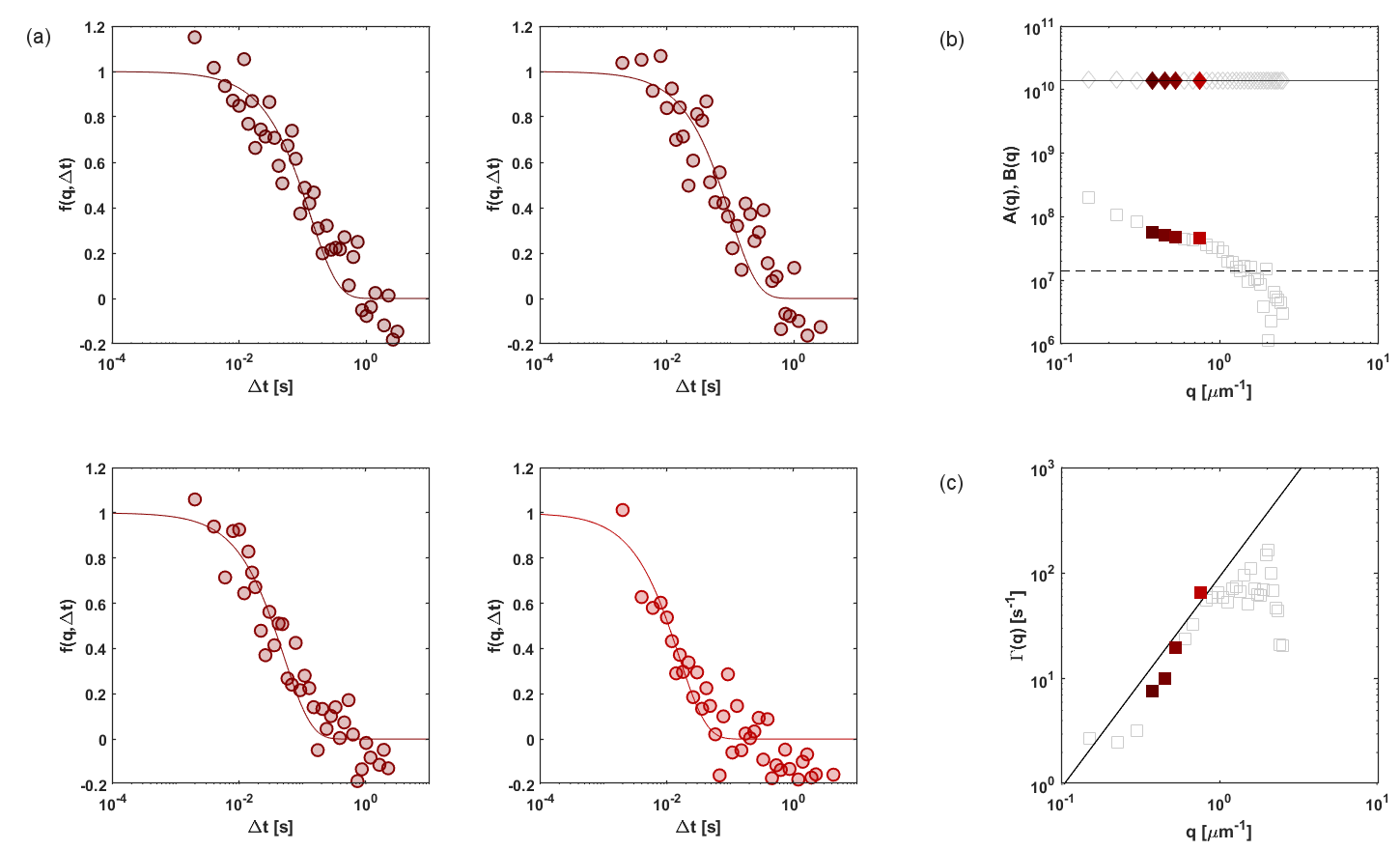}
\caption{\textbf{Lysozyme sample at $c=c_4$.} (a) Intermediate scattering functions $f(q,\Delta t)$ at different wavevectors $q$. From top left to bottom right: $q$=0.38 $\mu$m$^{-1}$, 0.45 $\mu$m$^{-1}$, 0.53 $\mu$m$^{-1}$, 0.76 $\mu$m$^{-1}$. Circles correspond to experimental data, solid lines correspond to a single exponential fit of the kind $f(q,\Delta t)=\exp(-\Gamma(q) \Delta t)$. (b) Signal amplitude $A(q)$ (squares) and noise term $B(q)$ (diamonds). The black solid line represents the average noise level, while the dashed line marks a SNR equal to $10^{-3}$. (c) Relaxation rates $\Gamma(q)$. Colored squares mark the range of $q$ considered for the quadratic fit. The weighted quadratic fit, shown as a black solid line, gives a diffusion coefficient $D$ equal to (92 $\pm$ 49) $\mu$m$^2$/s. Solid symbols in (b) and (c) highlight the wavevectors $q$ for which the decay of $f(q,\Delta t)$ is shown in panel (a).}
\label{fig:LYSO_C4}
\end{figure*}

\REV{For each protein solution we report in Table \ref{tab:ExpConc} the concentration values measured with the UV spectrophotometer, together with the extinction coefficients used for their calculation. The hydrodynamic radii reported in the literature for the proteins under study, as well as the values obtained from DLS measurements, are summarised in Table \ref{tab:ProteinSizeDLS}.
The corresponding protein size distributions are shown in Figure \ref{fig:DLS_dist}.
For each sample, we furthermore report in Figures \ref{fig:BSA_A1}-\ref{fig:LYSO_C4} examples of intermediate scattering functions at different wavevectors, as well as the amplitude, noise and relaxation rate values obtained from the fit.}\\

\begin{table}
\centering
\begin{tabular}{cccccc} 
 Protein & $c_1$ & $c_2$ & $c_3$ & $c_4$ & $\varepsilon$\\ 
 \hline
 BSA & 34 $\pm$ 2 & 11.7 $\pm$ 0.6 & 5.8 $\pm$ 0.3 & 1.17 $\pm$ 0.06 & 43824\\
 Pepsin & 23 $\pm$ 1 & 11.5 $\pm$ 0.6 & 3.9 $\pm$ 0.2 & 1.55 $\pm$ 0.08 & 52920\\
 Lysozyme & 28 $\pm$ 1 & 11.7 $\pm$ 0.6 & 8.3 $\pm$ 0.4 & 4.8 $\pm$ 0.2 & 36000\\ 
\end{tabular}
\caption{\textbf{Protein concentrations and extinction coefficients.} Concentration values are expressed in mg/mL. The extinction coefficients $\varepsilon$ are expressed in M$^{-1}$ cm$^{-1}$.}
\label{tab:ExpConc}
\end{table}

% measured values:
%\begin{table}
%\centering
%\begin{tabular}{cccccc} 
 %Protein & $c_1$ & $c_2$ & $c_3$ & $c_4$ & $\varepsilon$\\ 
 %\hline
 %BSA & 33.52 & 11.67 & 5.79 & 1.28 & 43824\\
 %Pepsin & 23.18 & 11.49 & 3.92 & 1.55 & 52920\\
 %Lysozyme & 28.37 & 11.72 & 8.27 & 4.80 & 36000\\ 
%\end{tabular}
%\caption{\textbf{Protein concentrations and extinction coefficients.} Concentration values are expressed in mg/mL. The extinction coefficients $\varepsilon$ are expressed in M$^{-1}$ cm$^{-1}$.}
%\label{tab:ExpConc}
%\end{table}

\begin{table}
\centering
\begin{tabular}{cccc} 
 Protein &  $R_h$ (literature) [nm] & $c$ & $R_h$ (DLS) [nm]\\ 
\hline
\multirow{4}{*}{BSA} & \multirow{4}{*}{3.3 - 4.3 \cite{jachimska2008characterization}} & $c_1$ &  3.03 $\pm$ 0.02\\% 3.0 - 3.5\\
%BSA & 3.3 - 4.3 & $c_1$ &  3.03 $\pm$ 0.02\\% 3.0 - 3.5\\
& & $c_2$ & 3.28 $\pm$ 0.01 \\
 & & $c_3$ & 3.37 $\pm$ 0.01 \\
 & & $c_4$ & 3.49 $\pm$ 0.01\\
 \hline
 \multirow{4}{*}{Pepsin} & \multirow{4}{*}{3.04 \cite{gtari2017impact}} & $c_1$ &  3.11 $\pm$ 0.03\\% 3.0 - 3.5\\
 & & $c_2$ & 3.29 $\pm$ 0.03 \\
 & & $c_3$ & 3.46 $\pm$ 0.07 \\
 & & $c_4$ & 3.51 $\pm$ 0.06\\
 \hline
 \multirow{4}{*}{Lysozyme} & \multirow{4}{*}{2.05 \cite{wilkins1999hydrodynamic}} & $c_1$ & 2.13 $\pm$ 0.01\\% 2.0 - 2.1\\
 & & $c_2$ & 2.04 $\pm$ 0.02 \\
 & & $c_3$ & 2.01 $\pm$ 0.01 \\
 & & $c_4$ & 1.96 $\pm$ 0.01\\
 \hline
\end{tabular}
\caption{\textbf{Protein size.} Mean hydrodynamic radii estimated with DLS for the three different proteins.}
\label{tab:ProteinSizeDLS}
\end{table}

\newpage

% BSA ----------------------------------------------

% sample A1

%\begin{figure*}[h]
%\centering
%\includegraphics[width=\textwidth]{Figures_SupportingInfo/BSA_A1_BIS.png}
%\caption{\textbf{Sample BSA A1. \CG{(BIS)}} (a) Intermediate scattering functions $f(q,\Delta t)$ at different wave vectors $q$. From top left to bottom right: $q$=0.45 $\mu$m$^{-1}$, 0.68 $\mu$m$^{-1}$, 0.98 $\mu$m$^{-1}$, 1.51 $\mu$m$^{-1}$. Circles correspond to experimental data, solid lines correspond to a single exponential fit of the kind $f(q,\Delta t)=\exp(-\Gamma(q) \Delta t)$. (b) Signal amplitude $A(q)$ (squares) and noise term $B(q)$ (diamonds). The horizontal solid line represents the average noise level $\bar{B}$, while the dashed line marks the limit on the SNR corresponding to $\bar{B}/300$. (c) Relaxation rates $\Gamma(q)$. Coloured squares mark the range of $q$ considered for the quadratic fit, in which the SNR is at least 1/300 and the relaxation rate does not exceed half the frame rate, namely $\Gamma(q)<250$ s$^{-1}$. The weighted quadratic fit, shown as a black solid line, gives a diffusion coefficient $D$ equal to (67 $\pm$ 2) $\mu$m$^2$/s. Solid symbols in both (b) and (c) highlight the wave vectors $q$ for which the decay of $f(q,\Delta t)$ is shown in panel (a).}
%\label{fig:BSA_A1}
%\end{figure*}

% sample A2

% sample A3

%\begin{figure*}[h]
%\centering
%\includegraphics[width=\textwidth]%{Figures_SupportingInfo/BSA_A4_LowSignal.png}
%\caption{\textbf{Sample BSA A4 \CG{(BIS)}.} (a) Intermediate scattering functions $f(q,\Delta t)$ at different wave vectors $q$. From top left to bottom right: $q$=0.38 $\mu$m$^{-1}$, 0.53 $\mu$m$^{-1}$, 0.60 $\mu$m$^{-1}$, 0.76 $\mu$m$^{-1}$. Circles correspond to experimental data, solid lines correspond to a single exponential fit of the kind $f(q,\Delta t)=\exp(-\Gamma(q) \Delta t)$. (b) Signal amplitude $A(q)$ (squares) and noise term $B(q)$ (diamonds). The horizontal solid line represents the average noise level $\bar{B}$, while the dashed line marks the limit on the SNR corresponding to $\bar{B}/350$. (c) Relaxation rates $\Gamma(q)$. Coloured squares mark the range of $q$ considered for the quadratic fit, in which the SNR is at least 1/350 and the relaxation rate does not exceed half the frame rate, namely $\Gamma(q)<250$ s$^{-1}$. The weighted quadratic fit, shown as a black solid line, gives a diffusion coefficient $D$ equal to (54 $\pm$ 20) $\mu$m$^2$/s. Solid symbols in both (b) and (c) highlight the wave vectors $q$ for which the decay of $f(q,\Delta t)$ is shown in panel (a).}
%\label{fig:BSA_A4_LowSignal}
%\end{figure*}

%%% PEPSIN ---------------------------

% sample B1

% sample B2

% sample B3

% sample B4

%%% LYSOZYME ---------------------------

% sample C1

% sample c2

% sample c3

% sample c4

\clearpage

\bibliography{bibl}